\documentclass[acmsmall,screen]{acmart}
\hypersetup{colorlinks=true, linkcolor=blue, urlcolor=blue}

\AtBeginDocument{%
  }

\usepackage[normalem]{ulem}
\usepackage{multirow}   
\usepackage{booktabs}   
\usepackage{pifont}     
\usepackage{array} 
\usepackage{ragged2e}
\usepackage{amsmath,amssymb,amsfonts}
\usepackage{graphicx}
\usepackage{textcomp}
\usepackage{makecell}
\usepackage{xcolor}
\usepackage{booktabs}
\usepackage{tikz}
\usepackage{ulem}
\usepackage{multirow}
\usepackage{pifont}
\usepackage{tcolorbox}
\usepackage{soul}
\usepackage{setspace}
\usepackage{threeparttable}
\usepackage{url}
\usepackage{subcaption}
\usepackage{graphicx}
\usepackage{makecell}
\usepackage{caption}
\usepackage{hyperref}
\usepackage[ruled,vlined,linesnumbered,boxed,commentsnumbered]{algorithm2e}

\makeatletter
\newcommand{\NoNumberLine}[1]{%
  \renewcommand{\@algocf@stepcounter}{\relax}%
  #1
  \renewcommand{\@algocf@stepcounter}{\refstepcounter}%
}
\makeatother
\usepackage{multirow}
\usepackage{algorithmicx}
\usepackage{algpseudocode} 
\usepackage{amsmath} 
\usepackage{graphicx}
\usepackage{float}
\usepackage{enumitem}

\usepackage{amsmath,amssymb,amsfonts}
\usepackage{algorithmicx}
\usepackage{graphicx}
\usepackage{textcomp}
\usepackage{makecell}
\usepackage{xcolor}
\usepackage{tikz}
\usepackage{ulem}
\usepackage{multirow}
\usepackage{pifont}
\usepackage{tcolorbox}
\usepackage{soul}
\usepackage{setspace}
\usepackage{threeparttable}
\usepackage{url}
\usepackage{subcaption}
\usepackage{graphicx}

\usepackage{titlesec}
\usepackage{graphicx}            
\usepackage{tikz}                
\usepackage[edges]{forest}       
\usepackage{xcolor}              
\usepackage{caption}             
\usepackage{float}               
\usepackage{epsfig}
\usepackage{enumitem}
\usepackage{fontawesome}

\usepackage{amsmath,amsfonts,bm}

\def\eqref#1{(\ref{#1})}

\def\1{\bm{1}}

\DeclareMathAlphabet{\mathsfit}{\encodingdefault}{\sfdefault}{m}{sl}
\SetMathAlphabet{\mathsfit}{bold}{\encodingdefault}{\sfdefault}{bx}{n}

\definecolor{sunye-red-light}{RGB}{235, 244, 255}
\definecolor{sunye-red-dark}{RGB}{114, 154, 202}

\definecolor{dingyifan-wangyixu-darkblue-light}{RGB}{245, 251, 246}
\definecolor{dingyifan-wangyixu-darkblue-dark}{RGB}{107, 182, 142}

\definecolor{wangxin-yellow-light}{RGB}{255, 254, 231}
\definecolor{wangxin-yellow-dark}{RGB}{238, 196, 84}

\usepackage{xspace}
\usepackage{arydshln}

\usepackage{xcolor}

\definecolor{citecolor}{RGB}{0, 0, 255}

\newcommand{\finding}[2]{
    \begin{center}
    \fcolorbox{black}{gray!10}{\parbox{.97\linewidth}{
    {#2}
    }}
    \end{center}
}

\setcopyright{acmcopyright}
\copyrightyear{2025}
\acmYear{2025}

\acmJournal{TOSEM}
\acmVolume{0}
\acmNumber{0}
\acmArticle{1}
\acmMonth{0}
\acmISBN{978-1-4503-XXXX-X/2018/06}

\begin{document}
\title{Understanding LLM-Centric Challenges for Deep Learning Frameworks: An Empirical Analysis}

\textcolor{blue}{\author{Yanzhou Mu}}
\orcid{0000-0003-1816-2246}
\affiliation{
  \institution{State Key Laboratory for Novel Software Technology, Nanjing University}
  \city{Nanjing}
  \country{China}
}
\email{602022320006@smail.nju.edu.cn}

\author{Rong Wang}
\orcid{0009-0004-0543-1894}
\affiliation{
  \institution{School of Information Science and Technology, Nantong University}
  \city{Nantong}
  \country{China}
}
\email{wangrong\_hcir@163.com}

\author{Juan Zhai}
\orcid{0000-0001-5017-8016}
\affiliation{
  \institution{University of Massachusetts Amherst}
  \city{Amherst}
  \state{MA}
  \country{USA}
}
\email{juanzhai@umass.edu}

\author{Chunrong Fang}
\orcid{0000-0002-9930-7111}
\affiliation{
  \institution{State Key Laboratory for Novel Software Technology, Nanjing University}
  \city{Nanjing}
  \country{China}
}
\email{fangchunrong@nju.edu.cn}
\authornote{Corresponding author.}

\author{Xiang Chen}
\orcid{0000-0002-1180-3891}
\affiliation{
  \institution{School of Artificial Intelligence and Computer Science, Nantong University}
  \city{Nantong}
  \country{China}
}
\email{xchencs@ntu.edu.cn}

\author{Jiacong Wu}
\orcid{0009-0006-5160-6418}
\affiliation{
  \institution{State Key Laboratory for Novel Software Technology, Nanjing University}
  \city{Nanjing}
  \country{China}
}
\email{776a6301@gmail.com}

\author{An Guo}
\orcid{0009-0005-8661-6133}
\affiliation{
  \institution{State Key Laboratory for Novel Software Technology, Nanjing University}
  \city{Nanjing}
  \country{China}
}
\email{guoan218@smail.nju.edu.cn}

\author{Jiawei Shen}
\orcid{0009-0007-0096-8221}
\affiliation{
  \institution{State Key Laboratory for Novel Software Technology, Nanjing University}
  \city{Nanjing}
  \country{China}
}
\email{221250056@smail.nju.edu.cn}

\author{Bingzhuo Li}
\orcid{0009-0001-2013-575X}
\affiliation{
  \institution{State Key Laboratory for Novel Software Technology, Nanjing University}
  \city{Nanjing}
  \country{China}
}
\email{231250105@smail.nju.edu.cn}

\author{Zhenyu Chen}
\orcid{0000-0002-9592-7022}
\affiliation{
  \institution{State Key Laboratory for Novel Software Technology, Nanjing University}
  \city{Nanjing}
  \country{China}
}
\email{zychen@nju.edu.cn}

\renewcommand{\shortauthors}{Mu, et al.}

\begin{abstract}
  Large language models (LLMs) have driven significant progress across a wide range of real-world applications. Realizing such models requires substantial system-level support. Deep learning (DL) frameworks provide this foundation by enabling efficient model construction, distributed execution, and optimized deployment. The large parameter scale and extended execution cycles impose exacting demands on deep learning frameworks, particularly in terms of scalability, stability, and efficiency. Therefore, poor usability, limited functionality, and subtle bugs in DL frameworks may hinder development efficiency and cause severe failures or resource waste. However, a fundamental question has not been thoroughly investigated in previous studies, i.e., \textbf{what challenges do DL frameworks face in supporting LLMs?} 
  To answer this question, we analyze issue reports from three major DL frameworks (i.e., MindSpore, PyTorch, and TensorFlow) and eight associated LLM toolkits such as Megatron. Based on a manual review of these reports, we construct a taxonomy that captures LLM-centric framework bugs, user requirements, and user questions. We then refine and enrich this taxonomy through interviews with 11 LLM users and eight DL framework developers. Based on the constructed taxonomy and findings summarized from interviews, our study further reveals key technical challenges and mismatches between LLM user needs and developer priorities. In summarization, our contributions are threefold: (1) we develop a comprehensive taxonomy comprising five question themes (11 sub-themes), five requirement themes (17 sub-themes), and ten bug themes (47 sub-themes); (2) we assess the importance and priority of different kinds of themes in the taxonomy based on LLM users and framework developers' insights; and (3) we identify five key findings across the LLM development and propose five actionable recommendations to improve the reliability, usability, and testability of DL frameworks. Our results highlight critical limitations in current DL frameworks and offer concrete guidance for advancing their support for the next generation of LLM construction and applications.
\end{abstract}



\received{20 February 2007}
\received[revised]{12 March 2009}
\received[accepted]{5 June 2009}

\maketitle

\section{Introduction}

Large language models (LLMs) have achieved widespread applications in various tasks, such as software engineering~\cite{wang2021codet5}, education~\cite{kasneci2023chatgpt}, healthcare~\cite{nazi2024large}, and finance~\cite{zhao2024revolutionizing}. They are composed of parameters ranging from billions to trillions, which require large amounts of storage and computing resources. To enable effective resource management and execution optimization during LLM training, inference, and deployment, DL frameworks offer support to provide distributed training, computational acceleration, and model quantization. For example, ChatGPT~\cite{openai2023gpt4}, one of the most widely used LLM applications, thoroughly relies on PyTorch, which delivers distributed parallelism for massive model parameters, GPU-optimized tensor operations, automatic differentiation for efficient optimization, and extensible APIs for custom attention mechanisms and memory management. Bugs in DL frameworks may significantly harm LLM systems, leading to severe economic losses and security incidents. For example, in June 2024, an attacker exploited a critical vulnerability~\cite{CVE-2024-3568} in the \textit{load\_repo\_checkpoint} interface of HuggingFace's Transformers library~\cite{huggingface2023models}, which serves as a foundational DL framework widely used for LLM loading and fine-tuning. The attacker injected malicious code (e.g., tampering with model weights), causing all LLMs loaded through this interface to silently fail during training. The attack remained undetected for weeks, resulting in massive GPU resource waste (e.g., over a month of training across thousands of nodes discarded) and estimated financial losses of over ten million dollars~\cite{motivatingexample}. \looseness=-1

Due to the scale and complexity of LLM workloads, LLM-centric framework bugs are challenging to detect yet may pose serious risks. They can lead to resource waste, extended execution time, or system crashes. 
Beyond bugs, the functionality completeness and usability of framework support critically influence user trust and offer essential references for developers to drive targeted improvements in supporting LLMs.
However, previous studies mainly focuses on designing testing methods~\cite{pham2019cradle,wang2020lemon,guo2020audee,li2023comet,Wei2022FreeLF,mu2024devmut,mu2025improving,deng2023large1,deng2023large2} for detecting bugs exposed in DL models with smaller parameter scale, like computer vision models that differ from LLMs in structural simplicity and parameter scale, or conducting empirical studies~\cite{chen2023toward, jia2021symptoms, tambon2024silent,hong2024investigating,zhang2020empirical,han2020programmers,wang2021automatic} on analyzing the symptoms, root causes, triggering conditions of common framework bugs, overlooking the complexity and sensitivity introduced by large language models. As LLMs adopt hybrid parallelism, modular fine-tuning, and low-bit inference, they place new demands on DL frameworks and expose novel bug patterns rarely observed in smaller models. 
Besides, the development and deployment of LLMs have significantly increased the difficulties for LLM users' coding and domain knowledge, requiring frameworks to provide users with convenient entry guides and user-friendly features to complete coding tasks. Overall, these challenges highlight the need to rethink defect classification and detection strategies tailored to the unique characteristics of LLMs. To our best knowledge, there is no research to investigate the challenges (e.g., framework bugs that harm LLMs, users' expectations regarding the support of LLMs by frameworks, and the priority of developers when dealing with LLM-centric issue reports) related to supporting LLMs of DL frameworks, which is critical for improving their quality.
Specifically, identifying missing features, usability gaps between users and frameworks, and frequently encountered bugs enables developers to prioritize upgrades, enhance framework quality, and attract a broader user base. Meanwhile, uncovering the obstacles developers encounter when diagnosing and resolving LLM-centric issues can help researchers design more effective testing and debugging techniques.\looseness=-1

Therefore, we conduct this empirical study to investigate users' expectations for LLM support in DL frameworks, common issues they often meet, bugs that most hinder their development, and the key obstacles and priorities developers face when addressing LLM-centric issues. Based on insights from both users and developers, we aim to uncover the most pressing challenges in enhancing LLM support within DL frameworks and propose potential directions for their optimization. Specifically, our study has two parts: \textbf{Stage I:} constructing an LLM-centric taxonomy of framework challenges from issue reports, and \textbf{Stage II:} enhancing the taxonomy via interviews with LLM users and framework developers. In \textbf{Stage I}, we collect and analyze issue reports from three major DL frameworks (i.e., PyTorch, MindSpore, and TensorFlow) and eight LLM toolkits (i.e., tools built on DL frameworks like PyTorch to streamline the development and deployment of LLMs such as Megatron for PyTorch). We filter and label LLM-centric cases by examining their titles, descriptions, and user–developer discussions to determine their basic type (e.g., bug, requirement, or question), general theme (e.g., root causes for bugs), and detailed content (e.g., bug symptoms).
Based on the labeled results, we construct an LLM-centric taxonomy of challenges in DL Frameworks that characterizes LLM-centric bugs, questions, and requirements consisting of three issue types (i.e., user question, requirement, and bug), 20 themes, and 75 sub-themes. In \textbf{Stage II}, we design an interview guideline based on the constructed taxonomy and interview eight framework developers who are specifically responsible for developing and testing LLM-supported functionality, and 11 LLM users who perform development, fine-tuning, or other purposes on LLMs to explore their development experiences, expectations for DL framework support, and views on current limitations.
We summarize key obstacles, expectations, and challenge priorities from their responses. Both developers and LLM users emphasize the correctness and stability of DL frameworks in supporting LLMs, often encountering distributed training failures, memory overflows, and numerical errors. LLM users further highlight gaps in configuration clarity, training observability, and reproducibility support. Developers concentrate on low-level concerns, including operator portability, memory alignment, and runtime robustness.  \looseness=-1




Based on the constructed taxonomy and interview results, we identify five key challenges that hinder DL frameworks from effectively supporting LLM development across different stages of the lifecycle. These challenges, perceived as most troublesome by LLM users or most critical by framework developers, range from environment setup to model deployment. Notably, two types of challenges emerge with particular urgency. First, environment setup remains fragile and error-prone due to dependency mismatches, undocumented hardware constraints, and incompatible toolchains, which frequently block newcomers and complicate reproducibility across platforms. Second, execution instability caused by silent training failures, non-deterministic behaviors, and misconfigured runtime states severely undermines model reliability and trust. Together, these challenges reflect deep-rooted mismatches between user expectations and framework capabilities in LLM workflows. Different from the general challenges related to bug pattern (i.e., symptoms, root causes, and trigger condition of bugs) with fixing and debugging concerns identified in prior studies~\cite{chen2023toward, jia2021symptoms, tambon2024silent, hong2024investigating, zhang2020empirical, han2020programmers, yan2025evaluating,wang2021automatic}, our identified challenges are specific to the scale, complexity, and modularity requirements of LLMs. To address them, we further propose five targeted optimization strategies aimed at improving framework readiness, developer experience, and support for scalable, production-grade LLM development. 

In summary, our key contributions are as follows:

\begin{itemize}

    \item \textbf{LLM-Centric Perspective.}
    We revisit the quality assurance of DL frameworks from the new perspective of LLM-centric challenges, addressing the growing demand for robust, scalable, and efficient frameworks driven by the widespread adoption of LLMs. 

    \item \textbf{Comprehensive Taxonomy.} 
    We analyze popular DL frameworks and their LLM toolkits, and construct a comprehensive taxonomy of questions, requirements, and bugs with 20 themes and 75 sub-themes that accurately reflect the gaps between LLM users and developers.

    \item \textbf{Empirical Validation through Expert Interviews.}
    We conduct in-depth interviews with real LLM users and framework developers to further validate the constructed taxonomy by adding missing challenges and collecting the priority of the items in the taxonomy from their perspectives. We further conclude five key findings that reflect the key challenges faced by LLM users and developers.

    \item \textbf{Actionable Suggestions.} Based on our findings, we further provide five optimization suggestions that cover the whole LLM development lifecycle, including environment installation, functionality support, execution stability, resource and performance optimization, and documentation enhancement.
    
\end{itemize}

\vspace{-5mm}
\section{Preliminary}

\subsection{Large Language Models} 
LLMs with hundreds of billions of parameters are trained on vast datasets for tasks like text classification, summarization, and generation. LLMs trace their origins to probabilistic methods like n-grams~\cite{suen1979n} and hidden Markov models~\cite{rabiner1989tutorial}, which struggled with long-distance dependencies and context. DL technologies like RNNs~\cite{lipton2015critical}, LSTMs~\cite{gers2000learning}, and especially transformers~\cite{vaswani2017attention}, greatly improved context understanding. Recent studies show that scaling transformer models to tens or hundreds of billions of parameters enhances context comprehension~\cite{zhao2023survey}. Transformers consist of (1) decoder-only models (e.g., GPT series~\cite{radford2018improving,radford2019language,brown2020language,openai2023gpt4}), (2) encoder-only models (e.g., BERT~\cite{devlin2018bert}), and (3) encoder-decoder hybrids (e.g., T5~\cite{raffel2020exploring}). 
Decoder-only models predict the next token based on prior context in an ``auto-regressive'' manner, excelling at tasks like text generation~\cite{becker2024text} and auto-completion~\cite{yang2019xlnet}. Encoder-only models mask words and predict them using context, performing well in tasks like question answering~\cite{chen2017reading} and text classification~\cite{rakhlin2016convolutional}. Encoder-decoder hybrids combine both approaches, making them ideal for text-to-text tasks such as machine translation~\cite{wu2016google} and summarization~\cite{mihalcea2004textrank,see2017get}.
In summary, decoder-only models excel in generation, encoder-only models in understanding, and hybrid models balance both tasks. Beyond text-focused models, multi-modal large language models (MLLMs) integrate data from different modalities like images, speech, and audio. MLLMs excel at cross-modal generation (e.g., text-to-image), retrieval (e.g., image-to-text), and applications in areas like video description, DNA structure prediction, and dialog generation. \looseness=-1

LLMs exhibit increasingly complex internal architectures, massive parameter scales, and diverse modality support (e.g., text-to-image). This growing complexity places substantially higher demands on DL frameworks, necessitating advanced capabilities such as efficient distributed training, large-scale memory management, heterogeneous hardware support, and flexible operator implementations. These heightened requirements, in turn, give rise to new challenges in ensuring the correctness, usability, and robustness of LLM support. Our study investigates these emerging challenges and understands how DL frameworks evolve to effectively meet the demands of modern LLM development.\looseness=-1

\subsection{ Deep Learning Frameworks in Supporting LLMs.} 
The development, execution, and deployment of LLMs heavily rely on DL frameworks, similar to small-scale DL models like VGG16 and ResNet50~\cite{he2016deep}. However, due to LLMs' strict efficiency and resource requirements, DL frameworks provide extreme optimizations that are not needed for smaller models. Techniques like quantization, pruning~\cite{han2015learning,tensorflow2019mot,molchanov2016pruning}, and distillation~\cite{gou2021knowledge} optimize LLMs, while mixed precision~\cite{micikevicius2017mixed} and layer fusion~\cite{kristiani2020optimization} accelerate model execution.
During development, DL frameworks offer standard modules (e.g., encoder~\cite{cho2014learning}, decoder~\cite{bahdanau2014neural}, embedding~\cite{pennington2014glove,bojanowski2017enriching}) for LLM architecture construction and efficient data processing. They optimize execution by pruning compute graphs and support data parallelism and optimizations like constant propagation and folding~\cite{cooper2022engineering}, as well as common sub-expression elimination~\cite{alfred2007compilers}.
In execution, DL frameworks manage hardware resources (e.g., CPUs, GPUs, NPUs) and streamline training and inference. They use parallelization techniques~\cite{jia2019beyond,dean2012large,hegde2016parallel}, including operator, pipeline, expert, and optimizer parallelism, to meet LLM demands. They also improve resource utilization through compute-communication fusion~\cite{li2014scaling} and memory reuse~\cite{gruslys2016memory}.
For deployment, DL frameworks provide conversion tools (e.g., MindSporeLite~\cite{MindSporeLite2020}) to optimize LLMs for specific environments (e.g., autonomous driving~\cite{ChenSKX15}), implementing hardware-specific strategies to enhance model efficiency. \looseness=-1

As LLM lifecycles depend on DL frameworks for architecture construction, execution optimization, and deployment adaptation, the scale and complexity of LLMs place unprecedented demands on framework functionality across all stages. Meeting these demands requires not only advanced performance optimizations and resource scheduling, but also seamless integration of heterogeneous hardware, robust distributed execution, and flexible deployment pipelines. These multifaceted requirements inevitably introduce new technical and usability challenges that differ from those in small-scale models. This study systematically investigates such challenges to understand how DL frameworks can be enhanced to better support the full lifecycle of modern LLMs.\looseness=-1

\begin{figure*}[htbp]
     \centering
    \includegraphics[width=0.95\textwidth]{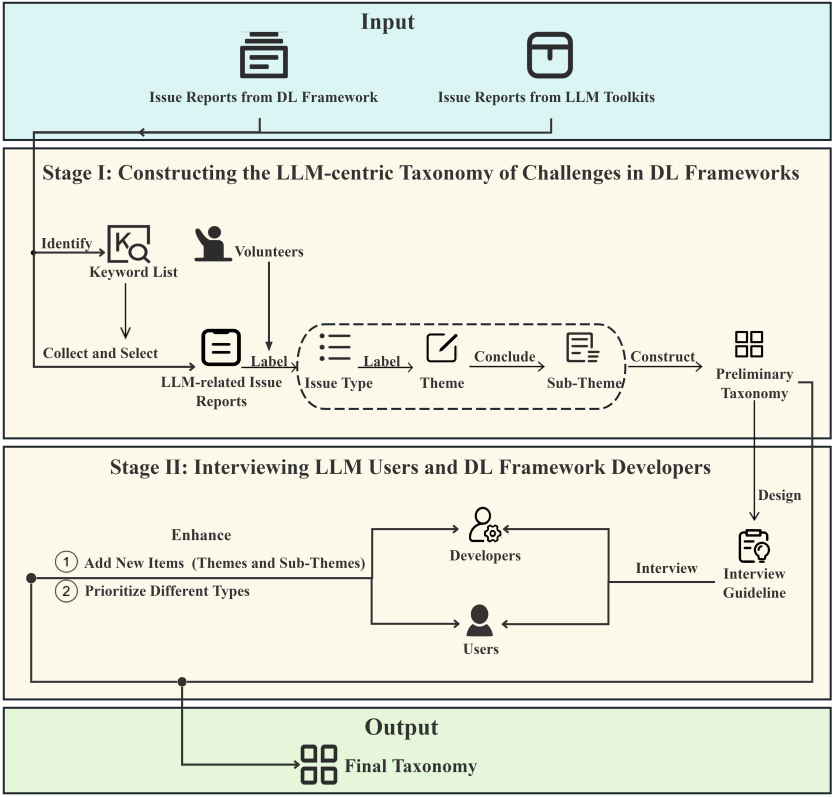}
    \vspace{-4mm}
     \caption{Workflow of Our Empirical Study}   
     
     \label{fig:workflow}
\end{figure*}

\vspace{-4mm}
\section{Research Motivation and Study Overview}
\subsection{Motivation of our Empirical Study}
As LLMs rapidly grow, the DL frameworks that support their development and deployment face increasing challenges in usability~\cite{narihira2021neural}, functionality~\cite{chen2020comprehensive}, and stability~\cite{haber2017stable}. The complexity and resource demands of LLMs amplify these challenges, reducing development efficiency and system reliability. However, previous work lacks a systematic analysis of the LLM-centric challenges in DL frameworks, and we show the motivations as follows.
\textbf{(1) High Demands from LLM Workloads.}
LLMs operate at a massive scale and require long execution cycles, pushing DL frameworks to deliver high scalability, stability, and efficiency. Even minor bugs can trigger major failures or waste costly resources, necessitating a deeper analysis and more effective testing strategies.
\textbf{(2) Fragmented Problem Landscape.}
Many GitHub maintainers report an increase in the volume of invalid AI-generated issues~\cite{register2024aislop}, which consumes developer time and disrupts maintenance workflows. This noise obscures real problems and reflects a lack of structure in current issue reporting. A clear taxonomy helps surface valid issues, reduce noise, and guide framework evolution. Yet, LLM-centric issues remain diverse and disorganized, making it hard to identify shared pain points or prioritize improvements.
\textbf{(3) Underexplored LLM users and Developer Perspectives.}
Previous studies~\cite{nguyen2019machine,han2020programmers,wang2021automatic} often ignore the lived experiences of users and framework developers. Their feedback reveals missing categories, clarifies vague terms, and highlights which issues matter most in practice, guiding more effective framework design and tool support. \looseness=-1

\subsection{Overview of Our Empirical Study} 
This study investigates challenges in DL frameworks that hinder users and developers from building or deploying LLMs and proposes optimizations to better support LLM workflows. As shown in Fig.~\ref{fig:workflow}, our study consists of two stages, i.e., (1) constructing the LLM-centric taxonomy for DL frameworks and (2) interviewing LLM users and DL framework developers.\looseness=-1

In Stage I, we investigate the following research question: \textbf{``RQ1: What types of LLM-centric challenges are exposed among DL frameworks and LLM toolkits?''} We aim to identify the common challenges that LLM users frequently encounter and report to framework communities, including common questions, requirements about new functionality, and bugs that threaten their LLM development. To answer this question, we collect LLM-centric issue reports from the communities of DL frameworks and LLM toolkits. We analyze issue titles, descriptions, and user–developer discussions to determine the type and content of each problem. Based on these analyses, we construct a taxonomy that categorizes the challenges into bugs, questions, and requirements. This taxonomy also serves as the foundation for designing the interview protocol in the next stage. The detailed setup and results are shown in Section~\ref{sec:taxonomy}.\looseness=-1

In Stage II, we further examine the taxonomy and insights obtained in Stage I by addressing the following research question: \textbf{``RQ2: How do LLM users and framework developers perceive and prioritize challenges from DL frameworks in real LLM development?''} To answer this question, we conduct interviews with LLM users and framework developers to gather their perspectives on framework support for LLMs. We refine the taxonomy with new categories that emerge from the interviews and identify which challenges are considered most critical in practice.The detailed setup and results are shown in Section~\ref{sec:interview}.\looseness=-1



\section{LLM-Centric Challenge Taxonomy for DL Frameworks}
\label{sec:taxonomy}
This section outlines the process and results of constructing our taxonomy, which progresses through three interconnected steps: (1) issue report collection, (2) manual labeling, and (3) bottom-up clustering.
The process begins by crawling issue reports from the communities of selected DL frameworks and their LLM toolkits. Because these repositories contain a large number of unrelated reports, we first compile an LLM-centric keyword list to extract potentially relevant issues. Volunteers validate this list to ensure precision and apply it to remove irrelevant reports. They then manually review the keyword-selected reports to eliminate false positives, such as keyword matches in usernames or file paths that automated filtering alone cannot catch. This two-stage screening addresses the challenge of reducing noise while preserving relevant cases for analysis.
The remaining issue reports often involve multiple topics, vague terminology, or implicit content that require careful contextual reasoning. Manual inspection is necessary to resolve these ambiguities, ensure semantic accuracy in category assignments, validate the boundaries between different top-level types and sub-themes, and provide the foundation for constructing the taxonomy. To achieve both quality and scalability, we recruit seven volunteers: six master’s students and one Ph.D. student, each with over a year of experience in LLM development and framework bug analysis for labeling and constructing the final taxonomy. All volunteers independently label the issue reports after filtering. They label each report as a type (e.g., Bug, Requirement, Question) and themes (e.g., root causes for bugs).
Finally, volunteers cluster the labeled reports using a bottom-up approach, merging similar themes into broader parent themes to form the final taxonomy. This step addresses the difficulty of organizing fragmented topics into a coherent structure. The resulting taxonomy follows a three-level hierarchy: (1) top-level types (Question, Requirement, Bug), (2) mid-level themes (e.g., (A.1) Unclear Installation Guidance), and (3) fine-grained sub-themes (e.g., (A.1.i) System Environment and Dependency Setup), enabling a detailed and scalable representation of LLM-centric challenges in DL frameworks, as illustrated in Fig.~\ref{fig:taxonomy}.\looseness=-1

\vspace{-4mm}
\subsection{Issue Reports Collection}
\label{sec:issuereportcollection}

Issue reports are user-submitted records, covering exceptions, feature requests, and questions that capture real-world development challenges, from functional bugs and performance bottlenecks to missing features and unclear documentation. For LLMs, DL frameworks and their toolkits serve as the primary environments for execution and development, making their community issue reports a direct reflection of the obstacles LLM users encounter in building, training, and deploying models. Analyzing these reports allows us to systematically identify LLM-specific challenges and inform targeted improvements in framework design. Therefore, we first select target DL frameworks and their LLM toolkits based on adoption scale, functional coverage, and relevance to real-world LLM development. 
For DL frameworks, we include PyTorch~\cite{pytorch}, TensorFlow~\cite{tensorflow}, and MindSpore~\cite{mindspore}. PyTorch and TensorFlow dominate global usage with mature ecosystems and active communities. MindSpore is a rapidly evolving framework that reveals challenges specific to NPU hardware and specialized deployment scenarios. For LLM toolkits, we examine DeepSpeed~\cite{deepspeed}, Megatron~\cite{megatron}, vLLM~\cite{vllm}, TensorRT-LLM~\cite{tensorrtllm}, ColossAI~\cite{colossalai}, MindSpeed~\cite{mindspeed}, MindNLP~\cite{mindnlp}, and MindFormers~\cite{mindformers}. These toolkits cover three core areas: (1) training acceleration and parallelism (e.g., DeepSpeed, Megatron, MindSpeed), (2) efficient inference and deployment (e.g., vLLM, TensorRT-LLM), and (3) model development and task abstraction (e.g., MindFormers, ColossAI, MindNLP). All selected projects show high community activity and are widely used in real LLM scenarios, providing a representative, practically grounded dataset that spans the entire lifecycle of training, fine-tuning, and inference.\looseness=-1

Then, issue reports are collected from their repositories on GitHub and Gitee between January 2023 and April 2025, since large numbers of developers and users have started developing and deploying LLMs after the release of ChatGPT~\cite{openai2023gpt4} at the end of 2022. 
Please note that the communities of general-purpose DL frameworks contain a vast number of issues unrelated to LLMs, and it is impractical to manually inspect them all. Therefore, we use a list of keywords to filter LLM-centric issue reports of basic DL frameworks, while all issues from LLM toolkits are collected directly. 
These keywords are derived from four complementary sources reflecting the ecosystem of LLM development: (1) model names and project identifiers commonly appearing in community discussions (e.g., llm, llama, baichuan, qwen, chatgpt, gemma, opensora, chatglm); (2) framework-specific APIs and modules widely used for large-scale training and inference (e.g., ``torch.distributed'', ``torch.compile'', ``tf.function'', ``graph mode''); (3) optimization and parallelization techniques essential for scaling LLMs (e.g., tensor parallel, pipeline parallel, auto\_parallel); and (4) hardware- and precision-related keywords indicative of deployment constraints (e.g., oom, cuda out of memory, bfloat16, int8, quantization, tpu, ascend). Keywords from these sources capture LLM-centric discussions from multiple perspectives, covering model-specific terminology, framework interfaces, optimization techniques, and hardware constraints, improving the accuracy of identifying relevant issue reports and filtering out noise, which is crucial for constructing a reliable dataset for downstream taxonomy development.
To address the risk of noise from ambiguous keywords that may introduce irrelevant issue reports, each candidate keyword is validated before adding it to the LLM-centric list. Three authors perform a two-step process combining LLM-generated summaries with human review, applied to 100 randomly sampled issues per keyword. First, authors use ChatGPT to generate a concise summary from the title and the first 300 words of the body, then classify the issue as (1) LLM-centric or (2) non-LLM-centric with a brief explanation. Next, two of the three authors independently verify each classification against the original content, and if they disagree, the remaining one of the three authors makes the final decision. Keywords whose precision is below 80\% are discarded. This process reduces manual workload while preserving expert judgment, ensures fairness and consistency through fixed-size sampling and precision thresholds, and produces a vetted keyword set that minimizes noise, improves dataset reliability, and supports reproducible large-scale studies of LLM-centric issues.
Finally, 72 keywords are selected for PyTorch, 84 for MindSpore, and 74 for TensorFlow, with full details available online~\cite{sharelink}. 
After constructing the keyword list, the three authors collect publicly available issues from the PyTorch, MindSpore, and TensorFlow communities and extract their titles and bodies. One of the three authors standardizes the text by converting it to lowercase and removing Markdown syntax and code blocks to reduce noise. When filtering LLM-centric issue reports,  the constructed list of LLM-centric keywords is used to match the issue content and retain issues containing at least one relevant term. This process produces a targeted subset of LLM-centric reports that supports subsequent analysis and research. After filtering by the keywords, there are 5,495, 3,285, and 3,568 LLM-centric issue reports left among PyTorch, MindSpore, and TensorFlow. \looseness=-1

After keyword filtering, some irrelevant issue reports remain because keywords may appear in non-LLM contexts, such as small-scale models using LLM modules, matches in usernames or file paths, or component names unrelated to LLM-specific behavior. To address this residual noise, three authors who validate the keywords as introduced above further perform a two-stage manual inspection to further reduce false positives. To balance quality and efficiency, they combine LLM-based pre-classification with stratified human verification.
In stage one, the three authors use ChatGPT to summarize each issue from its title and the first 300 words of the body, then assign it to one of three types: \textit{LLM-centric}, \textit{Non-LLM-centric}, or \textit{Unclear}. If summaries lack sufficient cues, they review the original report for clarification. In stage two, the three authors verify results using category-specific strategies: for \textit{LLM-centric} issues, they spot-check 10\% of samples and confirm a precision of 92.9\%, then retain the rest; for \textit{Unclear} cases, they review all entries; for \textit{Non-LLM-centric} cases, they sample 5\% and find a 1.2\% false negative rate, justifying the exclusion of the remainder.
During inspection, two of the three authors check whether an issue involves small-scale models with LLM modules but without LLM-specific topics, whether keywords appear only in usernames, file paths, or commits, and whether components bear LLM-centric names without actual LLM-related behavior. The remaining one of the three authors resolves disagreements when consensus cannot be reached.
The LLM-assisted summarization in the first stage can help rapidly screen large volumes of reports while preserving enough context for accurate judgment, and the stratified second stage focuses human review on the most uncertain cases, ensuring both efficiency and accuracy. This manual process for reducing false positives safeguards dataset quality for downstream taxonomy construction, improves precision by removing residual false positives, and reduces manual workload through targeted verification instead of exhaustive review.
After filtering, 38,742 valid LLM-centric reports are retained. Table~\ref{tab:issuereport} presents detailed statistics. The average false positive rate is 11.19\%.\looseness=-1

\begin{table}[]
  \centering
  \vspace{-4mm}
  \caption{Statistics of Issue Reports}
  \vspace{-4mm}
  \resizebox{1.02\textwidth}{!}{
    \begin{tabular}{c|ccc|ccc|c}
    \hline
    \multirow{2}[4]{*}{DL Frameworks and LLM ToolKit } & \multicolumn{3}{c|}{Issue Reports of Different Years} & \multicolumn{3}{c|}{Issue Reports of Different Types} & \multirow{2}[4]{*}{False Positive} \\
\cline{2-7}          & 2023  & 2024  & 2025  & Question & Bug   & Requirement &  \\
    \hline
    PyTorch & 1,593  & 1,573  & 1,015  & 15    & 3,997  & 169   & 23.93\% \\
TensorFlow & 908   & 1,167  & 349   & 9     & 2,335  & 80    & 32.12\% \\
MindSpore & 2,091  & 695   & 228   & 5     & 2,968  & 41    & 8.33\% \\
\cline{1-8}
DeepSeed & 2,100  & 1,380  & 270   & 51    & 2,551  & 1,148  & 5.15\% \\
Megatron & 373   & 572   & 169   & 120   & 639   & 355   & 8.60\% \\
vLLM & 1,220  & 8,423  & 3,903  & 130   & 7,459  & 5,957  & 9.74\% \\
TensorRT-LLM & 765   & 1,769  & 638   & 119   & 1,978  & 1,075  & 7.47\% \\
ColossAI & 2,838  & 940   & 77    & 21    & 1,619  & 2,215  & 14.35\% \\
MindNLP & 617   & 1,077  & 132   & 7     & 720   & 1,099  & 7.83\% \\
MindSpeed & 0     & 68    & 12    & 3     & 52    & 25    & 2.50\% \\
MindFormers & 698   & 854   & 228   & 62    & 1,391  & 327   & 3.09\% \\
\hline
Total   & 13,203 & 18,518 & 7,021  & 542   & 25,709 & 12,491 & Average: 11.19\% \\

    \hline
    \end{tabular}%
    }
  \label{tab:issuereport}%
\end{table}%

\newcommand{\TLM}{LLM-centric taxonomy of Challenges in DL Frameworks}

\definecolor{sunye-red-light}{RGB}{235,244,255}
\definecolor{sunye-red-dark}{RGB}{114,154,202}
\definecolor{dingyifan-wangyixu-darkblue-light}{RGB}{245,251,246}
\definecolor{dingyifan-wangyixu-darkblue-dark}{RGB}{107,182,142}
\definecolor{wangxin-yellow-light}{RGB}{255,254,231}
\definecolor{wangxin-yellow-dark}{RGB}{238,196,84}

\tikzstyle{vfm-bg-node}=[rectangle, rounded corners=2pt,
  fill=sunye-red-light, draw=sunye-red-dark,
  inner sep=3pt, line width=1pt,
  font=\tiny, align=left]
\tikzstyle{vfm-line-node}=[rectangle, rounded corners=2pt,
  draw=sunye-red-dark, fill opacity=.2,
  inner sep=3pt, line width=0.5pt,
  font=\tiny, align=left]
\tikzstyle{llm-bg-node}=[rectangle, rounded corners=2pt,
  fill=dingyifan-wangyixu-darkblue-light, draw=dingyifan-wangyixu-darkblue-dark,
  inner sep=3pt, line width=1pt,
  font=\tiny, align=left]
\tikzstyle{llm-line-node}=[rectangle, rounded corners=2pt,
  draw=dingyifan-wangyixu-darkblue-dark, fill opacity=.2,
  inner sep=3pt, line width=0.5pt,
  font=\tiny, align=left]
\tikzstyle{vlp-bg-node}=[rectangle, rounded corners=2pt,
  fill=wangxin-yellow-light, draw=wangxin-yellow-dark,
  inner sep=3pt, line width=1pt,
  font=\tiny, align=left]
\tikzstyle{vlp-line-node}=[rectangle, rounded corners=2pt,
  draw=wangxin-yellow-dark, fill opacity=.2,
  inner sep=3pt, line width=0.5pt,
  font=\tiny, align=left]

\begin{figure*}[]
  \centering
  \resizebox{0.90\textwidth}{!}{%
    \begin{forest}
      forked edges,
      for tree={
        grow=east,
        reversed,
        anchor=mid west,
        parent anchor=mid east,
        child anchor=mid west,
        rectangle,
        draw=none,
        align=left,
        font=\tiny,
        edge+={darkgray,line width=0.6pt},
        s sep=2.8pt, inner xsep=1.8pt, inner ysep=2.7pt,
        minimum width=2em,
        par/.style={rotate=90, child anchor=north, parent anchor=south, anchor=center},
      },
      where level=0{font=\footnotesize}{},
      where level=1{text width=3.8em}{},
      where level=2{text width=13.8em}{},
      [ {\TLM}, par
        [ Question, vfm-bg-node, text width=5em
          [ (A.1) Unclear Installation Guidance, vfm-line-node, text width=16em
            [ {\textbf{(i) System Environment and Dependency Setup}\\
               \textbf{(ii) Runtime Configuration and Launch Parameters} }, vfm-line-node, text width=17em]
          ]
          [ (A.2) Usage Gaps in LLM Operations, vfm-line-node, text width=16em
            [ {\textbf{(i) User Support and Documentation Quality}\\
               \textbf{(ii) Functionality Support and Boundary Clarification} }, vfm-line-node , text width=17em]
          ]
          [ (A.3) Unexpected Behavior and Unintuitive Design, vfm-line-node, text width=16em
            [ {\textbf{(i) Framework Implementation and Design Issues}\\
               \textbf{(ii) Resource Efficiency and Optimization Management}\\
               \textbf{(iii) Model Behavior and Output Verification} }, vfm-line-node, text width=17em ]
          ]
          [ (A.4) Cross-Platform Deployment Issues, vfm-line-node, text width=16em
            [ {\textbf{(i) Engineering Implementation and Deployment Challenges}\\
               \textbf{(ii) Environmental Compatibility Challenges} }, vfm-line-node, text width=17em ]
          ]
          [ (A.5) Training Strategy and Configuration Confusion, vfm-line-node, text width=16em
            [ {\textbf{(i) Confusion over complex training strategies}\\
               \textbf{(ii) Strategy failure after version upgrades} }, vfm-line-node, text width=17em ]
          ]
        ]
        [ Requirement, llm-bg-node, text width=5em
          [ (B.1) New Environment Compatibility and Deployment, llm-line-node, text width=16em
            [ {\textbf{(i) Hardware Adaptation and Compatibility Requirements}\\
               \textbf{(ii) Deployment and Toolchain Optimization}\\
               \textbf{(iii) Dependency Version and Compatibility Management} \\
               \textbf{(iv) Local Development and Lightweight Setup Support}
               }, llm-line-node , text width=17em]
          ]
          [ (B.2) System and Model Feature Support, llm-line-node, text width=16em
            [ {\textbf{(i) Modular API and Interface Support	}\\
               \textbf{(ii) Scalable Distributed Training Support}\\
               \textbf{(iii) Large-scale Data Ingestion and Preprocessing	}\\
               \textbf{(iv) Model Expansion and Format Interoperability}}, llm-line-node , text width=17em]
          ]
          [ (B.3) Code Reliability on supporting LLM of DL Frameworks, llm-line-node, text width=16em
            [ {\textbf{(i) Code Quality and Architecture Optimization}\\
               \textbf{(ii) Test Automation and Quality Assurance}\\
               \textbf{(iii) Debugging and Logging Requirements}\\
               \textbf{(iv) Exception Handling and Fault-Tolerance} }, llm-line-node, text width=17em ]
          ]
          [ (B.4) Community and Document Support, llm-line-node, text width=16em
            [ {\textbf{(i) Documentation and Tutorial Support}\\
               \textbf{(ii) Collaboration Workflows and Toolchain Support} \\
               \textbf{(iii) Minimal Working Examples
               }}, llm-line-node , text width=17em]
          ]
          [ (B.5) Training Workflow and Efficiency Optimization, llm-line-node, text width=16em
            [ {\textbf{(i) Runtime Performance and Resource Management }\\
               \textbf{(ii) Training Configuration and Scalability Support captures} }, llm-line-node , text width=17em]
          ]
        ]
        [ Bug, vlp-bg-node, text width=5em
          [ (C.1) System Compatibility and Integration Failures, vlp-line-node, text width=16em
            [ {\textbf{(i) Build, Installation, and Low-Level Support Failures}\\
               \textbf{(ii) Operator Compilation and Backend Support Failures}\\
               \textbf{(iii) Versioning and Compatibility Breakages}\\
               \textbf{(iv) Environment Dependency and Toolchain Compatibility Bugs}\\
               \textbf{(v) Hardware and Accelerator Adaptation Failures}\\
               \textbf{(vi) Platform and Deployment Bugs} }, vlp-line-node , text width=17em]
          ]
          [ (C.2) Resource Efficiency and Memory Management, vlp-line-node, text width=16em
            [ {\textbf{(i) Scheduling and Parameter Management Bugs}\\
               \textbf{(ii) Performance Optimization and Monitoring Failures}\\
               \textbf{(iii) Quantization Accuracy and Inference Performance Bugs}\\
               \textbf{(iv) Memory Fragmentation and Leak}
               }, vlp-line-node , text width=17em]
          ]
          [ (C.3) Distributed and Parallel Execution, vlp-line-node, text width=16em
            [ {\textbf{(i) Distributed Communication and Initialization Bugs}\\
               \textbf{(ii) Distributed/Multi-Process Data Communication Failures}\\
               \textbf{(iii) Communication and Coordination Errors}\\
               \textbf{(iv) Hybrid Parallelism Misconfiguration}
               }, vlp-line-node , text width=17em]
          ]
          [ (C.4) Numerical Stability and Precision Management, vlp-line-node, text width=16em
            [ {\textbf{(i) Numerical Computation and Operator Stability Bugs}\\
               \textbf{(ii) Precision and Quantization Error Management Bugs}\\
               \textbf{(iii) Checkpoint Precision Compatibility Failures} }, vlp-line-node, text width=17em ]
          ]
          [ (C.5) Training Strategy and Execution Control, vlp-line-node, text width=16em
            [ {\textbf{(i) Training Strategy and Optimizer Usage Bugs}\\
               \textbf{(ii) Model Training Stability Failures}\\
               \textbf{(iii) Initialization and State Configuration Anomalies}\\
               \textbf{(iv) Optimizer and Gradient Strategy Errors}\\
               \textbf{(v) Model Unresponsiveness or Erroneous Inference Outputs} }, vlp-line-node , text width=17em]
          ]
          [ (C.6) Data Pipeline and I/O Management, vlp-line-node, text width=16em
            [ {\textbf{(i) Data Flow Processing Failures}\\
               \textbf{(ii) Data Reading and Input Errors}\\
               \textbf{(iii) Data Preprocessing and Type Handling Bugs}\\
               \textbf{(iv) Model Persistence and Saving Failures}\\
               \textbf{(v) Model File and Checkpoint Compatibility Bugs} }, vlp-line-node, text width=17em ]
          ]
          [ (C.7) Interface Design and Configuration Management, vlp-line-node, text width=16em
            [ {\textbf{(i) API Interface and Parameter Configuration Errors}\\
               \textbf{(ii) Interface Specification and Compatibility Bugs}\\
               \textbf{(iii) Interface Tool and Configuration Management Failures}\\
               \textbf{(iv) Configuration Parameter Handling Anomalies}\\
               \textbf{(v) Configuration Logging and Documentation Consistency Bugs} }, vlp-line-node, text width=17em ]
          ]
          [ (C.8) Framework Module Implementation and Integration, vlp-line-node, text width=16em
            [ {\textbf{(i) Module Integration and Compatibility Errors}\\
               \textbf{(ii) Core Framework Mechanism Bugs}\\
               \textbf{(iii) Architectural Module Execution Failures}\\
               \textbf{(iv) Module Design Logic Bugs}\\
               \textbf{(v) Inference and Generation Anomalies} }, vlp-line-node, text width=17em ]
          ]
          [ {(C.9) Code Quality, Logging, Testing and Security}, vlp-line-node, text width=16em
            [ {\textbf{(i) Testing and Quality Assurance Bugs}\\
               \textbf{(ii) Code and Algorithm Implementation Failures}\\
               \textbf{(iii) Error Handling and Boundary Condition Bugs}\\
               \textbf{(iv) Logging, Debugging, and Monitoring Bugs}\\
               \textbf{(v) Runtime Errors}\\
               \textbf{(vi) Security and Stability Failures} }, vlp-line-node, text width=17em ]
          ]
          [ (C.10) Documentation and Maintainability, vlp-line-node, text width=16em
            [ {\textbf{(i) Documentation and Standards Gaps}\\
               \textbf{(ii) Documentation Metadata and Annotation Bugs}\\
               \textbf{(iii) Documentation Toolchain and Configuration Sync Failures}\\
               \textbf{(iv) User Interaction and Documentation Errors} }, vlp-line-node , text width=17em]
          ]
        ]
      ]
    \end{forest}%
  }
  \vspace{-4mm}
  \caption{LLM-centric Taxonomy of Challenges in DL Frameworks}
  \label{fig:taxonomy}
\end{figure*}
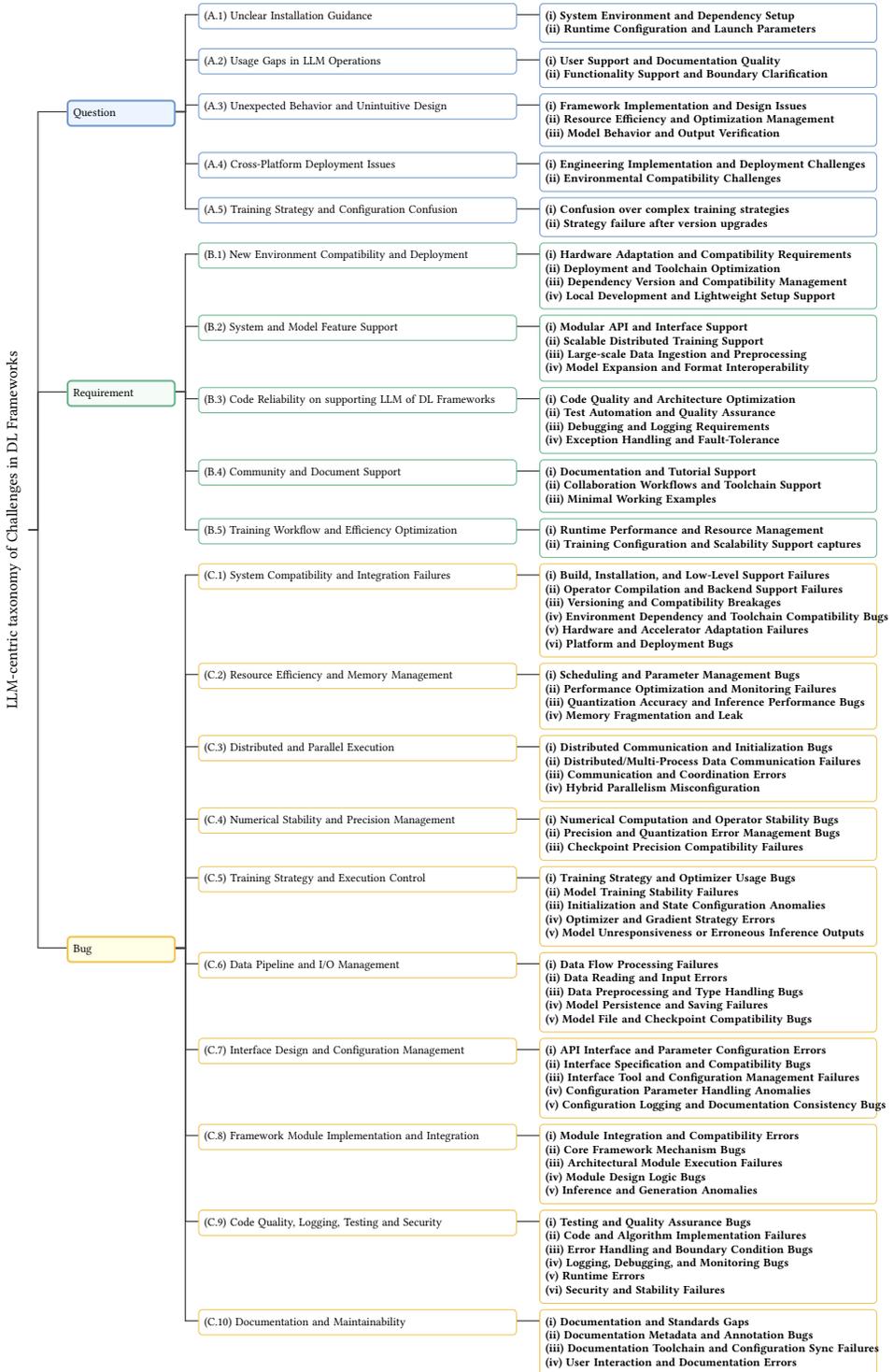

\subsection{Issue Reports Labeling}
\label{sec:issuereportlabel}
As introduced at the beginning of Section~\ref{sec:taxonomy}, we recruit seven volunteers, including one Ph.D. student and six master’s students, each with over one year of experience in LLM development and handling framework bugs across various DL frameworks, to label the issue reports after keyword filtering. They assign each report a type, theme, and sub-theme. All volunteers receive training before labeling. A Ph.D. student serves as the lead annotator and conducts a pilot study on 7,000 randomly sampled reports (18.07\% of all selected issue reports), following established practices~\cite{zimmermann2007predicting,hindle2008automatic,sillito2020failures} that typically analyze about 20\% of data to extract features, identify patterns, and summarize topics. The large scale and diversity of LLM-centric reports make this pilot study essential for establishing a consistent labeling framework before full-scale annotation, as differences in interpretation could lead to inconsistent labels, blurred category boundaries, and costly rework. This analysis yields three primary issue types:\looseness=-1

\begin{itemize}
\item \textbf{Question}: LLM user confusion regarding framework usage, configuration, or supported functionalities.
\item \textbf{Requirement}: Requests for new features, functionality enhancements, or compatibility improvements.
\item \textbf{Bug}: Reports of deviations from expected behavior, such as execution errors or incorrect results.
\end{itemize}

The lead annotator determines each issue’s type using the ChatGPT-generated summaries from the false positive filtering stage, combined with the title and description, and then identifies core themes by analyzing comments and keywords (e.g., out-of-memory, optimization, precision loss). For each issue, the lead annotator writes a brief summary of its symptoms or suggested fixes, which aids understanding but is not part of the taxonomy.

After the pilot study, the lead annotator presents the tag set and annotation process to all volunteers in an online meeting. The group reviews tag definitions, discusses examples, and resolves ambiguities. Disagreements are resolved collaboratively until consensus is reached. This pilot study ensures that all volunteers share a consistent understanding of the labeling rules, reducing misclassification and rework, improving efficiency during large-scale labeling, and establishing a standardized process that future annotators can follow, thereby enhancing transparency and reproducibility.

All seven volunteers then independently annotate the remaining reports, following a consistent process: (1) read the GPT-generated summary to quickly understand the issue; (2) if the summary is insufficient, review the original title, description, comments, logs, and code snippets; (3) determine the issue type (question, requirement, or bug), assign a theme (e.g., root causes for bugs, gaps for requirements, or topics for questions), and provide supporting details. To reduce subjectivity, each volunteer labels the full set of reports. New categories are added only when all volunteers unanimously agree. Compared with the initial tags from the pilot study, only five new tags are added: A.4.i for questions, B.3.iv and B.1.iii for requirements, and C.10.ii and C.10.iii for bugs (see Fig.~\ref{fig:taxonomy}), confirming that the pilot study captures the majority of categories.

After the initial round, volunteers jointly review all annotations and re-label controversial cases, defined as reports receiving an equal number of different labels. Disagreements are discussed in an online meeting, with the lead annotator making the final decision if consensus cannot be reached. This process, combining independent full-set annotation, consensus-based category expansion, and joint review, ensures label consistency and prevents arbitrary tag proliferation. The high Fleiss’ $Kappa$ scores, i.e., 0.717 for bugs, 0.755 for requirements, and 0.757 for questions, indicating strong inter-annotator agreement, confirming the dataset’s reliability and reproducibility as a foundation for downstream taxonomy construction and analysis.
\looseness=-1

\subsection{Taxonomy Constrction}
After labeling, all volunteers apply a bottom-up approach~\cite{vijayaraghavan2003bug} to construct the taxonomy of framework-level issue reports in LLMs, covering bug, question, and requirement reports. This process organizes the large number of fine-grained themes from the labeling stage into a coherent, hierarchical structure that captures the breadth and depth of LLM-related challenges since it can avoid overlapping semantics and inconsistent boundaries between themes would hinder meaningful analysis and interpretation.
Each volunteer first independently clusters similar themes, i.e., root causes for bug reports, motivations for requirement reports, and topics for question reports. They then abstract a parent theme for each cluster to represent the common characteristics of its sub-themes, ensuring each sub-theme maintains an ``is-a'' relationship with its parent.
After the initial grouping, all volunteers hold an online meeting to unify naming conventions, clarify definitions, and determine the coverage scope of each sub-theme. They also verify that every sub-theme is correctly assigned to its parent theme. Finally, all volunteers collectively review the complete taxonomy, including all issue types, themes, and sub-themes, discussing and resolving controversial cases to ensure consistent categorization.
The resulting taxonomy captures a comprehensive set of LLM-centric challenges across bugs, questions, and requirements. Once finalized, volunteers classify explanatory details for each sub-theme (e.g., symptoms for bug-related sub-themes, specific functionalities for requirement-related sub-themes, and detailed questions for question-related sub-themes) using the summarized and detailed information from the labeling results. These explanations are stored separately from the taxonomy hierarchy to maintain structural clarity while preserving interpretability and traceability for later reference.
This process transforms dispersed labeling results into a structured, scalable taxonomy, ensures consistent categorization through collaborative review, and produces a framework that future studies can directly reuse or extend.\looseness=-1

\subsection{Details of the Final Taxonomy}
\label{sec:rq1result}

\begin{figure*}[]
    \centering
    \begin{subfigure}[b]{0.30\textwidth}
        \raisebox{0.5em}{\includegraphics[width=\textwidth]{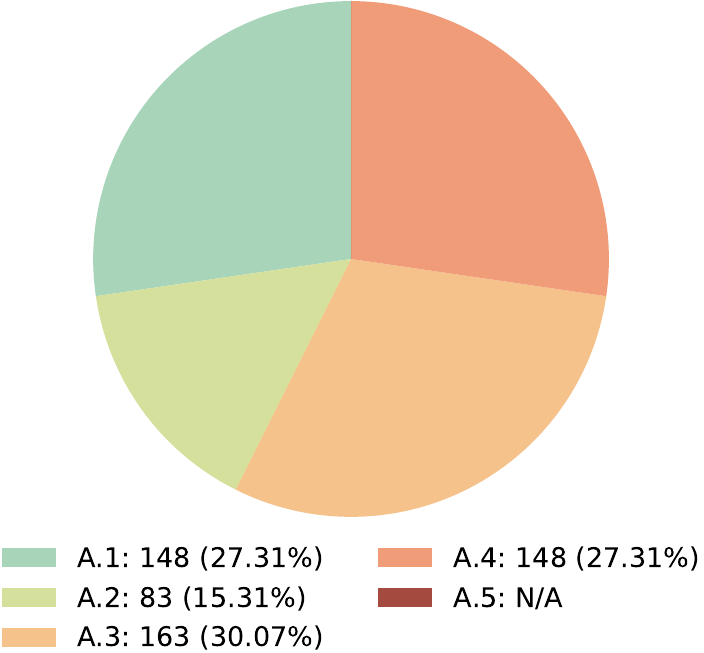}}
        \vspace{-4mm}
        \caption{Question}
        \label{fig:Question}
    \end{subfigure}
    \begin{subfigure}[b]{0.33\textwidth}
        \raisebox{0em}{\includegraphics[width=\textwidth]{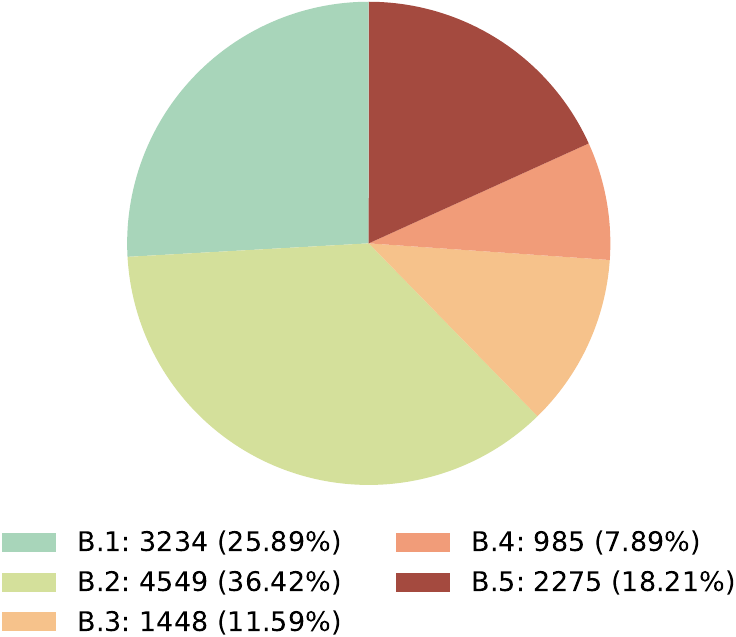}}
        \vspace{-4mm}
        \caption{Requirement}
        \label{fig:Requirement}
    \end{subfigure}
    \begin{subfigure}[b]{0.33\textwidth}
        \raisebox{0.1em}{\includegraphics[width=\textwidth]{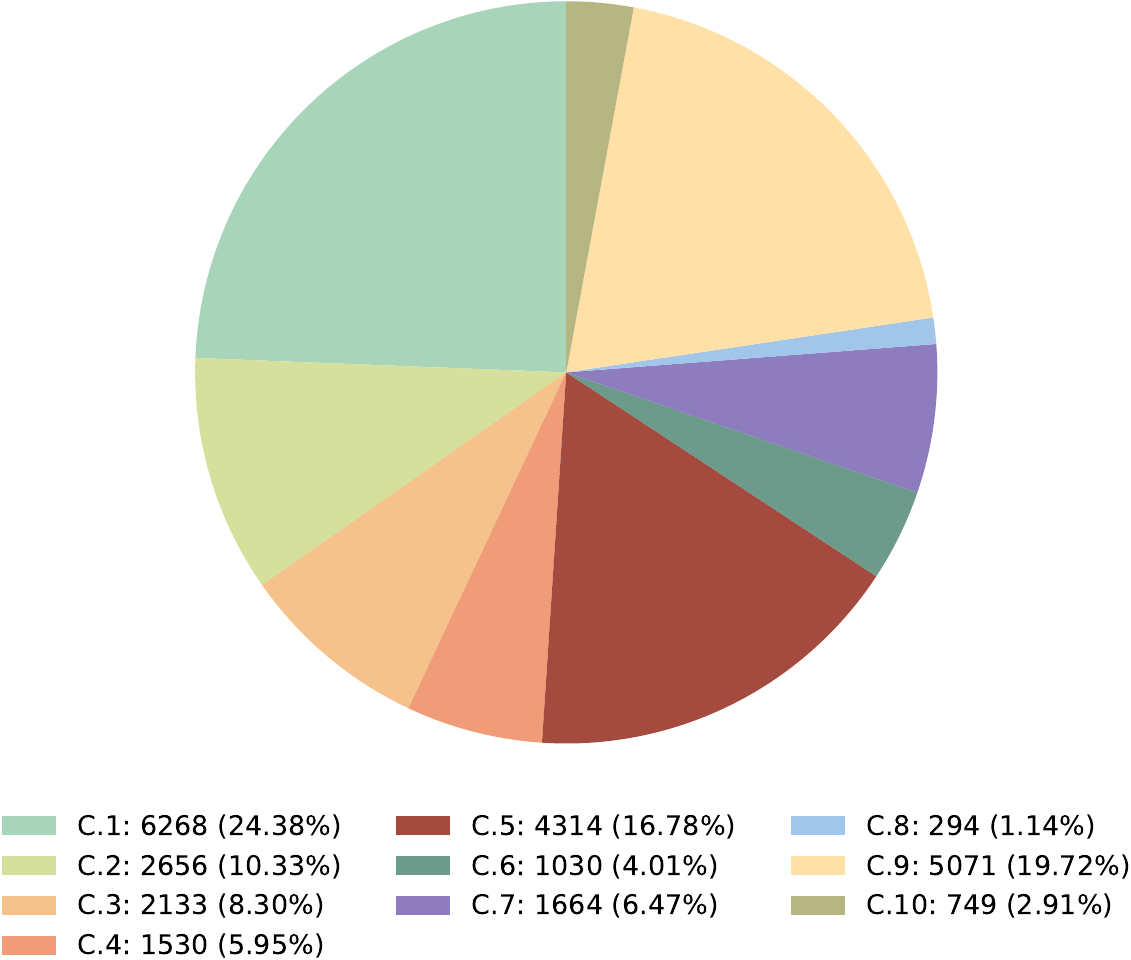}}
        \vspace{-4mm}
        \caption{Bug}
        \label{fig:Bug}
    \end{subfigure}
   \vspace{-4mm}
    \caption{Statistics of Different Themes of our Taxonomy \footnotesize  (Notice that A.5 is a newly added theme from interviews and thus is not included in the proportion statistics.) }

    \label{fig:taxonomydistributed}
\end{figure*}

The constructed taxonomy organizes all annotated issue reports into three top-level types.
The question type comprises five themes, 11 sub-themes, and 297 distinct question intents.
The requirement type comprises five themes, 17 sub-themes, and 754 distinct feature expectations.
The bug type comprises ten themes, 47 sub-themes, and 1,586 unique bug symptoms.
Fig.~\ref{fig:taxonomydistributed} presents the distribution of reports across the three types and different themes.
To ensure practical relevance, the taxonomy is further refined through interviews with LLM users and DL framework developers. This led to the addition of new themes and sub-themes, the splitting of several existing ones, and the renaming of others.
Table~\ref{tab:interviewmodify} summarizes these modifications, while the final, complete taxonomy is described in detail in the following sections.\looseness=-1

\begin{table*}[]
  \centering
  \vspace{-4mm}
  \caption{Modifications from Interview Results on the Taxonomy}
  \vspace{-4mm}
  \resizebox{1.02\textwidth}{!}{
    \begin{tabular}{l|l}
    \hline
    Suggestions & Details \\
    \hline
    \multirow{5}[2]{*}{Add New Themes/Sub-Themes} & Add new Theme A.5 and relevant Sub-Themes A.5.i and A.5.ii. \\
          & Add new Sub-Theme B.1.iv \\
          & Add new Sub-Theme B.4.iii \\
          & Add new Sub-Theme C.2.iv \\
          & Add new Sub-Theme C.3.iv \\
    \hline
    Clarify the Meaning of existing Themes/Sub-Themes & clarify the boundary between A.2  and A.3. \\
   \hline
    \multirow{4}[2]{*}{Modify the Title of Themes/Sub-Themes} & Rename A.1.i from ``System Support and Dependency Management'' to ``System Environment and Dependency Setup'' \\
          & Rename A.1.ii from ``Configuration and Parameter Usage'' to Runtime Configuration and Launch Parameters. \\
          & Rename C.5 from ``Training and Inference Control Logic'' to ``Training Strategy and Execution Control''. \\
          & Rename C.8 from ``Architecture and Module Integration'' to ``Framework Module Implementation and Integration''. \\
    \hline
    Change Existing Themes/Sub-Themes & Split B.2 into two categories: the refined B.2 (System and Model Feature Support) and add new B.5 (Training Workflow and Efficiency Optimization).  \\
    \hline
    \end{tabular}%
    }
  \label{tab:interviewmodify}%
\end{table*}%



\subsubsection{Question Type\\}
Such issue reports expose persistent knowledge gaps in LLM development. We summarize key sub-themes as follows.

\begin{figure*}[htbp]
     \centering
    \includegraphics[width=0.9\textwidth]{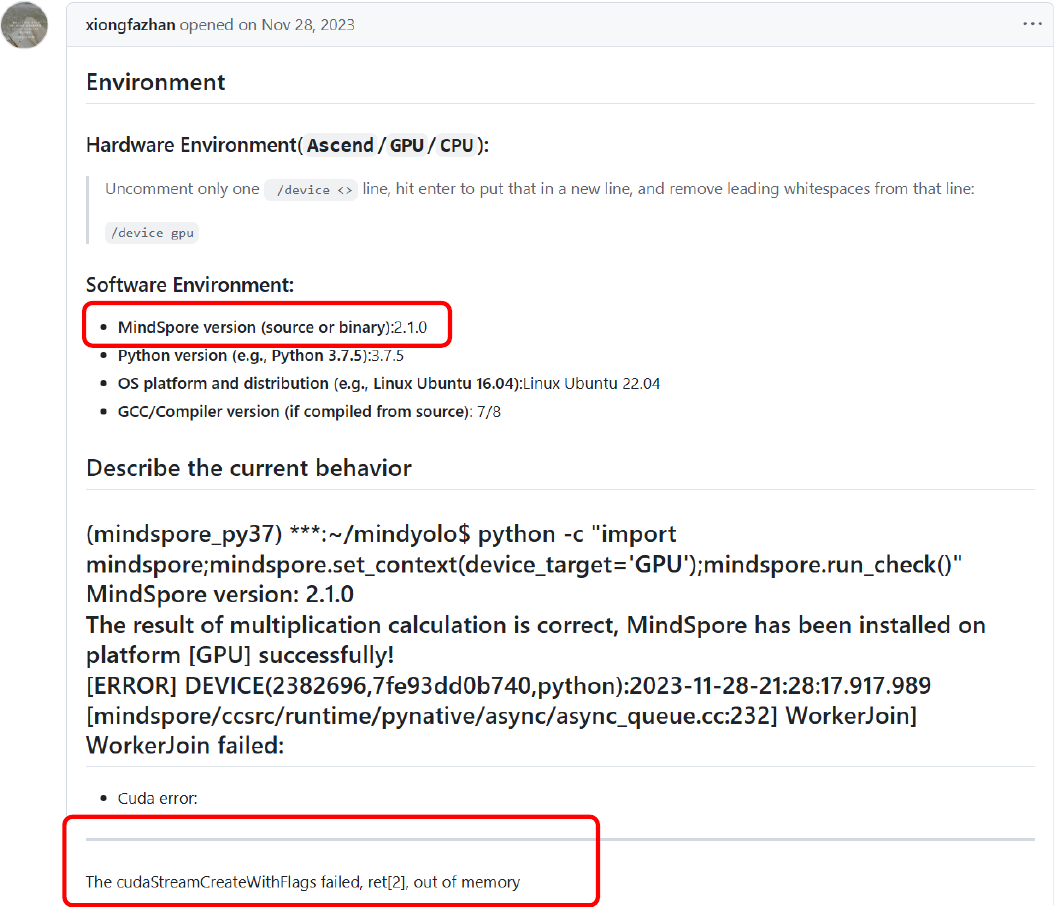}
    \vspace{-4mm}
     \caption{A sample post in Model Behavior and Output Verification (A.1.ii)} 
     \vspace{-2mm}
     \label{fig:A12}
\end{figure*}

\textbf{\faQuestionCircle\ 1.Unclear Installation Guidance (A.1).}
\label{taxonomy:a.1}
This theme covers 148 reports (27.31\% of all question-related issues) involving setup problems that block framework initialization or execution. LLM users often struggle with missing documentation, unclear dependency requirements, and undocumented default behaviors. 
\textit{System Support and Dependency Management (A.1.i)} involves environment-level failures before successful installation, typically due to mismatched CUDA/cuDNN versions, missing GPU libraries, or broken build toolchains. For example, one LLM users failed to build with CUDA~12.1 on the latest PyTorch Docker image~\cite{a1case1}.  
\textit{Configuration and Parameter Usage (A.1.ii)} captures post-installation misconfigurations, such as incorrect device targets, misuse of mixed precision flags, or invalid launch parameters. For instance, one LLM user experienced GPU malfunction on Ubuntu~22.04 after installing version 2.1.0 due to an unsupported runtime configuration~\cite{a1case3}, as shown in Fig.~\ref{fig:A12}.  
A.1.i focuses on environment-level incompatibilities requiring system fixes, whereas A.1.ii addresses user-level errors that block execution. Both undermine usability, especially for newcomers without system expertise. \looseness=-1

\textbf{\faQuestionCircle\ 2.Usage Gaps in LLM Operations (A.2).}
\label{taxonomy:a.2}
This theme includes 83 reports, accounting for 15.31\% of all questions. It covers obstacles users face when interacting with LLM frameworks due to unfamiliar operational workflows, vague documentation, and under-specified features. Users frequently express confusion about how to use, configure, or understand various features, parameters, and behaviors of LLM frameworks. These gaps easily cause silent errors or wasted GPU resources, especially in complex LLM settings.
\textit{Documentation Gaps and Configuration Confusion (A.2.i)} involves difficulties in applying core LLM features such as precision modes, optimizer settings, and parallel training strategies. For example, users report uncertainty about initiating multi-GPU training in Megatron~\cite{a2case1} or confirming support for mixed-precision training~\cite{a2case2}.  
\textit{Ambiguous Module Boundaries and Behavioral Inconsistencies (A.2.ii)} concern unclear module capabilities and their interactions with model variants. For instance, a user fails to apply vLLM to the LLaMA interface and asks which models support the \texttt{limit\_cm\_persprompt} parameter~\cite{a2case3}.  
A.2.i reflects insufficient guidance for common configuration tasks, whereas A.2.ii exposes implicit assumptions and ill-defined feature boundaries. Both underscore the need for clearer documentation, explicit feature specifications, and tooling aligned with users’ mental models of LLM workflows.\looseness=-1

\begin{figure*}[htbp]
     \centering
    \includegraphics[width=0.9\textwidth]{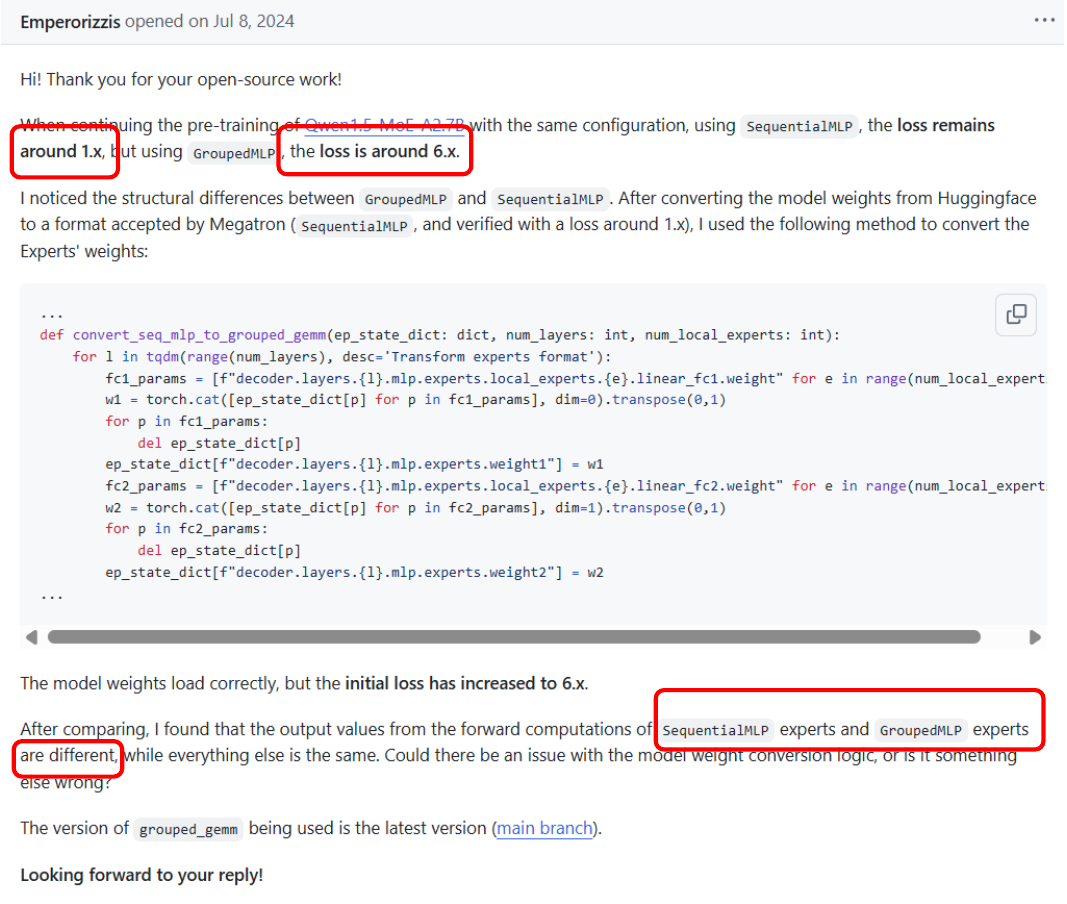}
    \vspace{-4mm}
     \caption{A sample post in Model Behavior and Output Verification (A.3.iii)}
     \vspace{-2mm}
     \label{fig:A33}
\end{figure*}

\textbf{\faQuestionCircle\ 3.Unexpected Behavior and Unintuitive Design (A.3).}
\label{taxonomy:a.3}
This theme comprises 163 reports, accounting for 30.07\% of all questions. It captures user confusion resulting from system behaviors that appear abnormal, inconsistent, or unintuitive during training or inference. Users often encounter abnormal training behaviors, performance issues, and inconsistent outputs that deviate from expectations, without clear explanations from the framework to help diagnose the cause.
\textit{Confusing Framework Design (A.3.i)} includes unclear design choices or poorly documented mechanisms that obscure expected behaviors. For example, users question how to validate that learning rate schedulers in PyTorch behave as intended~\cite{pytorch-issue22107}.
\textit{Resource Management Issues (A.3.ii)} cover challenges such as unexpected out-of-memory (OOM) errors in TensorFlow when resource allocation fails silently~\cite{tensorflow-issue46111}.
\textit{Model Behavior and Output Verification (A.3.iii)} involves difficulty in interpreting training outcomes, particularly when users observe abnormal loss values, repeated outputs, or unstable gradients. One Megatron user, for instance, reports erratic loss behavior when training MoE models with more than eight experts per group~\cite{megatron-issue909}, as shown in Fig.~\ref{fig:A33}. 
These questions often arise from missing diagnostic tooling (e.g., sanity checks, validation hooks, internal assertions) and the lack of behavioral baselines across setups. Without verification mechanisms or reference benchmarks, users cannot easily tell whether anomalies stem from framework bugs, misconfigurations, or inherent training variance.  Compared with A.2, which asks ``\textit{How should I do this?}'' about unclear usage, A.3 asks ``\textit{Why is this happening?},'' revealing deeper gaps in observability, correctness, and design clarity. \looseness=-1

\textbf{\faQuestionCircle\ 4.Cross-Platform Deployment Issues (A.4).}
\label{taxonomy:a.4}
This theme comprises 148 reports (27.31\% of all question-related issues) on deployment challenges when transferring trained models across hardware platforms or toolchains. Users often ask whether specific models, features, or configurations work across different hardware, environments, or frameworks, and frequently encounter compatibility errors that block execution, cause deployment failures, or require repeated environment adjustments.  
\textit{System Architecture Incompatibility (A.4.i)} covers failures from differences in hardware architectures, operating systems, runtime libraries, or device drivers (1.05\% of all questions). Users report degraded performance or persistent errors when migrating models from NVIDIA GPUs to CPUs, Apple Silicon, TPUs, or custom accelerators, often due to unsupported operators, precision mismatches, or inconsistent runtime behaviors.  
\textit{Export Compatibility and Tooling Gaps (A.4.ii)} addresses issues in model serialization, checkpoint conversion, and inference export tools (0.69\% of all questions). Examples include difficulties converting checkpoints between PyTorch and TensorRT~\cite{a4case3}, deployment failures on Apple Silicon~\cite{a4case1}, and loss of numerical precision when moving from GPU to CPU~\cite{a4case2}.  
These two sub-themes reflect distinct but complementary challenges: A.4.i concerns hardware-level inconsistencies that undermine runtime correctness, while A.4.ii involves tooling limitations that disrupt model portability. In production settings, both pose significant risks to efficiency, stability, and deployment success. \looseness=-1

\textbf{\faQuestionCircle\ 5.Training Strategy and Configuration Confusion (A.5).}
This theme emerges from interview feedback and was absent in the initial volunteer-based labeling, so no quantitative statistics are available. Unlike previous categories, these questions arise not from framework defects but from limited understanding of configuration interactions, training strategies, and behavior changes after version updates. Without clear documentation and given varying expertise levels, users encounter unstable training, degraded performance, or unexplained results, prompting them to seek guidance from developers or experienced LLM users.  
\textit{Confusion over Complex Training Strategies (A.5.i)} covers difficulties in configuring hybrid parallelism, freezing layers, or coordinating optimizers and schedulers. Missteps often cause ineffective gradient updates, diverging loss, or non-functional learning rate schedules.  
\textit{Strategy Failure after Version Upgrades (A.5.ii)} addresses cases where previously working configurations fail after framework or toolkit updates, leaving users unsure whether issues stem from undocumented behavior changes, altered defaults, or version incompatibilities. \looseness=-1

\textbf{Summary.} User questions, accounting for 1.40\% of all reports, reveal persistent knowledge gaps and usability barriers across the LLM development pipeline. The largest share (30.07\%, A.3) concerns abnormal runtime behaviors, such as unstable loss, repeated outputs, or non-converging gradients, without sufficient tools to judge whether they are expected or indicate deeper faults. Installation and configuration issues (27.31\%, A.1) involve CUDA compatibility, cuDNN mismatches, and vague setup guidance. Deployment to alternative platforms like Apple Silicon or ROCm (27.31\%, A.4) presents additional challenges, including hardware incompatibilities and tooling gaps. Unclear documentation or feature boundaries (15.31\%, A.2) often lead to silent errors or wasted resources, while strategy failures after version upgrades (A.5), identified qualitatively through interviews, disrupt established workflows despite previously functional configurations. While A.2 and A.3 both reflect user confusion, A.2 centers on ``\textit{How should I do this?}'' questions about unclear usage, whereas A.3 centers on ``\textit{Why is this happening?}'' questions about unexpected behaviors. Together, these reports underscore a fundamental misalignment between framework complexity and users’ capacity to observe, configure, and reason about LLM workflows, particularly in early development stages. \looseness=-1

\finding{1}{
 \textbf{Finding 1.} 
   User questions reveal a persistent mismatch between framework design assumptions and the knowledge boundaries of LLM users. Installation and configuration confusion accounts for 27.31\% of reports (A.1), while unexpected runtime behaviors without clear diagnostic cues form the largest share at 30.07\% (A.3). These issues show that users are often blocked not by missing features but by insufficient guidance, vague configuration instructions, and the lack of explicit failure signals such as meaningful error messages or validation hooks. Concentrated in the early stages of environment setup and debugging, these challenges hinder progress and increase trial-and-error costs.\looseness=-1

}

\subsubsection{Requirement Type\\} 

LLM-centric requirement reports reflect user expectations for DL frameworks, comprising 32.24\% of all entries in our taxonomy. We summarize the themes and sub-themes below. \looseness=-1

\begin{figure*}[htbp]
     \centering
    \includegraphics[width=0.8\textwidth]{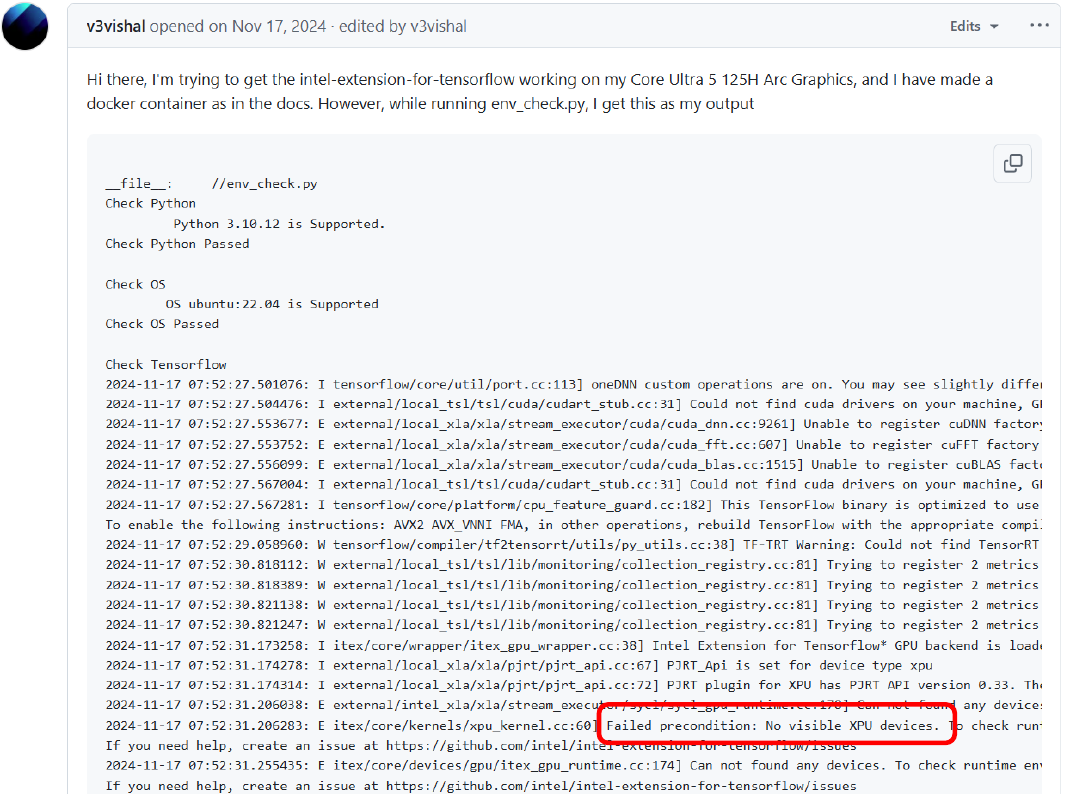}
    \vspace{-4mm}
     \caption{A sample post in Deployment and Toolchain Optimization (B.1.ii)}   
     \label{fig:B12}
\end{figure*}

\textbf{\faLightbulbO\ 1.New Environment Compatibility and Deployment (B.1).}
\label{taxonomy:b.1}
This theme includes 3,234 reports (25.89\% of all requirement-related issues) capturing user demands for broader hardware support, smoother deployment, and lightweight local development. Users request compatibility with new or non-standard hardware, operating systems, and toolchains.  
\textit{Hardware Adaptation (B.1.i)} covers support for emerging accelerators (e.g., XPUs, HPUs), alternative OSs (e.g., Windows, ROCm), and CPU-only modes~\cite{b1case1}.  
\textit{Deployment and Toolchain Optimization (B.1.ii)} involves reducing manual setup in containerized environments like Docker, with requests for clearer workflows, pre-configured images, or simplified orchestration tools~\cite{b1case2}, as shown in Fig.~\ref{fig:B12}.  
\textit{Dependency Compatibility (B.1.iii)} reflects frustration with ecosystem fragility, where updates to core packages (e.g., NumPy) frequently break downstream dependencies and require manual fixes~\cite{b1case3}.  
\textit{Local Development and Lightweight Setup Support (B.1.iv)} addresses CPU-only, Windows/WSL2, or non-Docker environments, with requests for verified CPU runtimes, conda-based installations, and quick-start examples enabling prototyping without full GPU infrastructure.  
Collectively, these sub-themes reveal a gap between framework deployment assumptions and real-world environments. Users call for installation paths that are minimal, reproducible, and accessible, especially for local debugging, experimentation, and early-stage development.\looseness=-1

\textbf{\faLightbulbO\ 2.Functionality Support for LLM Development (B.2).}
\label{taxonomy:b.2}
This theme includes 4,549 reports (36.42\% of all requirement-related issues) capturing user expectations for enhanced functionality across the LLM development pipeline. Users request missing APIs, broader model compatibility, flexible configuration interfaces, and functional improvements essential for building, loading, or operating LLMs. Without these capabilities, many abandon the framework at the design or prototyping stage.  
\textit{Modular API and Interface Support (B.2.i)} covers demands for modularized interfaces, fine-grained I/O control, and integrated preprocessing utilities to streamline experimentation and system integration~\cite{b2case1}.  
\textit{Scalable Distributed Training Support (B.2.ii)} reflects the need for robust multi-GPU and multi-node training with better orchestration, fault tolerance, and elasticity~\cite{b2case2}.  
\textit{Large-scale Data Ingestion and Preprocessing (B.2.iii)} highlights requirements for efficient handling of large, heterogeneous datasets, including streaming, preprocessing pipelines, and format adaptation~\cite{b2case4}.  
\textit{Model Expansion and Format Interoperability (B.2.iv)} involves extending existing models, resuming from checkpoints, and ensuring compatibility with third-party formats and toolkits~\cite{b2case5}.  
Collectively, these sub-themes emphasize the need for end-to-end functionality that scales with model size, training infrastructure, and diverse data and model ecosystems.\looseness=-1

\textbf{\faLightbulbO\ 3.Code Reliability in LLM Support (B.3).}
\label{taxonomy:b.3}
This theme includes 1,448 reports (11.59\% of all requirement-related issues) reflecting user expectations for robust, maintainable, and fault-tolerant frameworks. Users seek improvements in code structure, testing, logging, and exception handling to prevent crashes, silent failures, and incorrect outputs that are difficult to trace or fix.  
\textit{Code Quality and Architecture (B.3.i)} covers requests for modular, well-structured codebases that ensure consistent behavior and support long-term maintenance~\cite{b3case1}.  
\textit{Test Automation (B.3.ii)} emphasizes comprehensive unit and integration testing within CI pipelines to detect regressions and compatibility issues early~\cite{b3case2}. 
\textit{Debugging and Logging (B.3.iii)} involves deadlock detection, runtime state tracking, and informative logging for diagnosing and recovering from large-scale training issues~\cite{b3case3}.  
\textit{Exception Handling (B.3.iv)} addresses robust failure recovery through retry mechanisms, checkpoint restoration, and consistent cross-component error propagation~\cite{b3case4}.  
Together, these demands emphasize the importance of engineering rigor in LLM frameworks, where reliability, observability, and automated validation play a key role in enabling stable development at scale.\looseness=-1

\textbf{\faLightbulbO\ 4.Community and Documentation Support (B.4).}
\label{taxonomy:b.4}
This theme includes 985 reports (7.89\% of all requirement-related issues) reflecting expectations for accessible learning resources, effective collaboration, and practical onboarding. Users request better documentation, clearer usage examples, and standardized contribution processes.  
\textit{Documentation and Tutorial Support (B.4.i)} covers demands for accurate, up-to-date API references and deployment guides, such as configuring ZeRO-3 for scalable training~\cite{REQUEST}.  
\textit{Collaboration Workflows (B.4.ii)} involves clear issue templates, contributor guidelines, and responsive maintainer feedback, as seen in TensorRT-LLM’s RFC-based design discussions~\cite{tensorrt_llm_rfc_feedback}.  
\textit{Minimal Working Examples (B.4.iii)} highlights the need for clean, runnable, and reproducible templates that demonstrate key functionalities with valid configurations and verified compatibility, covering tasks like single-GPU fine-tuning, distributed inference, and checkpoint recovery.  
Collectively, these sub-themes highlight the importance of high-quality documentation and community infrastructure in reducing onboarding costs and enabling sustainable engagement across diverse user groups.\looseness=-1

\begin{figure*}[htbp]
     \centering
    \includegraphics[width=0.8\textwidth]{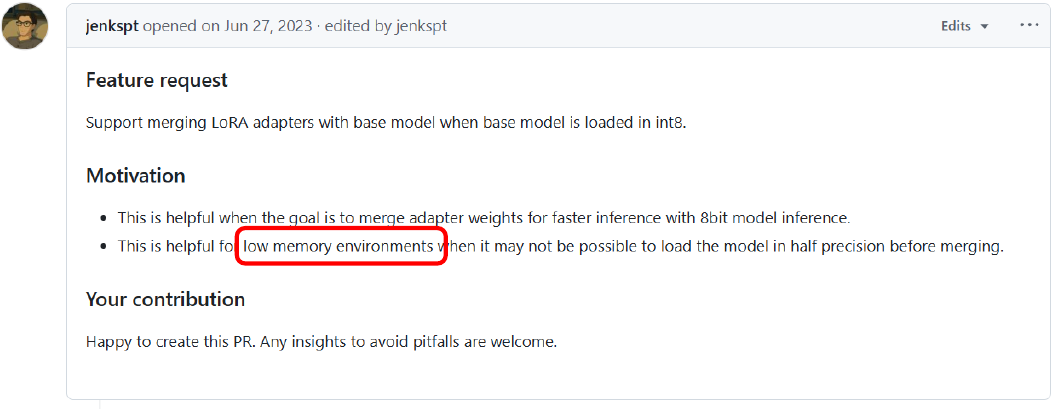}
    \vspace{-4mm}
     \caption{A sample post in Training Configuration and Scalability Support (B.5.ii)}   
     \vspace{-4mm}
     \label{fig:B52}
\end{figure*}

\textbf{\faLightbulbO\ 5.Training Workflow and Efficiency Optimization (B.5).} 
\label{taxonomy:b.5}
This theme includes 2,275 reports (18.21\% of all requirement-related issues) reflecting demand for efficient, configurable training workflows under increasing computational constraints. Users propose enhancements for training efficiency, parallelism strategies, optimizer state handling, checkpointing, and memory/resource management to support large-scale LLM training. Without these capabilities, models can be built but not trained effectively, limiting real-world applicability.  
\textit{Runtime Performance and Resource Management (B.5.i)} covers improvements to training throughput, memory efficiency, and fault tolerance. Users request better handling of GPU memory fragmentation, mitigation of large-batch slowdowns, and fine-grained performance monitoring via trace visualizations and profiling tools~\cite{b2case3}.  
\textit{Training Configuration and Scalability Support (B.5.ii)} focuses on flexibility and automation in training setups, including long-context input, large-batch training with gradient accumulation, mixed-precision and quantization-aware execution, and configuration auto-tuning for scalability~\cite{b2case6}, as shown in Fig.~\ref{fig:B52}.  
Collectively, these sub-themes emphasize the need for tunable, performance-aware training pipelines that remain stable and efficient at LLM scale.\looseness=-1

\textbf{Summary.} Requirement reports account for 32.24\% of all entries, showing that user expectations extend far beyond baseline functionality. The largest share (36.42\%, B.2) concerns functional scalability, including modular APIs, distributed training support, large-scale data ingestion pipelines, and model extensibility. Deployment flexibility (25.89\%, B.1) follows, with demands for CPU-only setups, emerging accelerators (e.g., XPUs, HPUs), and lightweight local environments. Engineering quality is another recurring theme: 11.59\% of reports (B.3) request more robust architecture, automated testing, and logging infrastructure. Although smaller in proportion (7.89\%, B.4), community and documentation support are critical for onboarding and long-term engagement. Performance tuning and workflow efficiency (18.21\%, B.5) target stable, resource-aware execution at LLM scale. While B.2 focuses on delivering complete capabilities, B.5 emphasizes optimizing those capabilities for efficiency and scalability; similarly, B.1 addresses environment readiness, whereas B.4 focuses on collaborative resources for using those environments effectively. Collectively, these patterns show that users view frameworks as end-to-end platforms requiring robustness, transparency, and engineering maturity.\looseness=-1

\finding{1}{
 \textbf{Finding 2.} 
     Requirement reports indicate a shift in user expectations from frameworks as functional toolkits to engineering platforms. Over one-third (36.42\%, B.2) request enhanced LLM development capabilities, while 25.89\% (B.1) emphasize deployment flexibility across diverse hardware and environments. Additional demands for engineering quality (B.3), community and documentation support (B.4), and performance-aware training workflows (B.5) reinforce that users now prioritize not only functionality but also scalable integration, operational robustness, and infrastructure maturity.\looseness=-1
}

\begin{figure*}[htbp]
     \centering
    \includegraphics[width=0.8\textwidth]{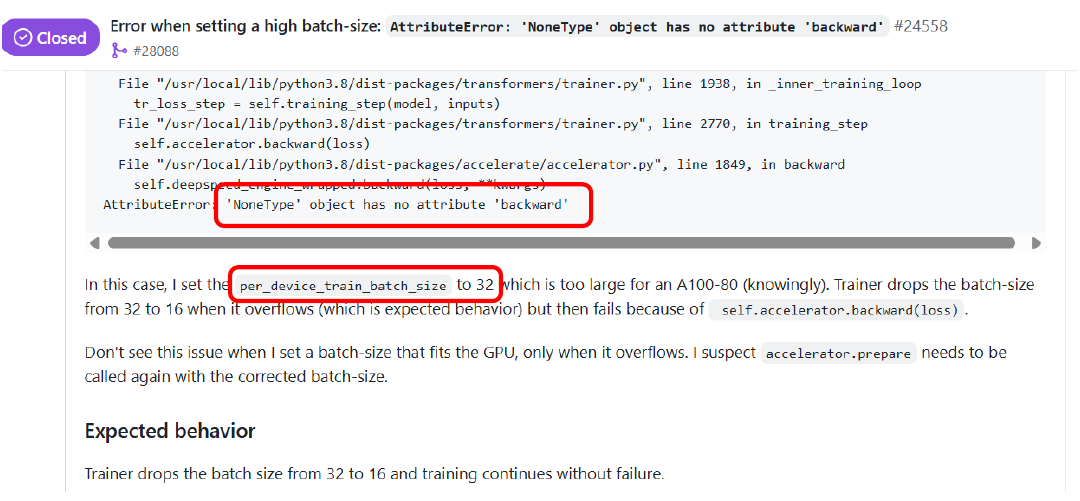}
    \vspace{-4mm}
     \caption{A sample post in Scheduling and Parameter Management Errors (C.2.i)}   
     \label{fig:C21}
\end{figure*}

\subsubsection{Bug Type\\}

LLM-centric bug reports account for 66.36\% of all entries in our taxonomy, reflecting key obstacles users face in scaling models, tuning workloads, and deploying systems. We summarize the key themes as follows. \looseness=-1

\textbf{\faBug\ 1.System Compatibility and Integration Failures (C.1).}
\label{taxonomy:c.1}
This theme includes 6,268 reports (18.21\% of all bug-related issues) arising from mismatches between frameworks and target system environments, which prevent models from being built, initialized, or executed. These failures often occur during model building, installation, or execution due to platform, environment, or dependency incompatibilities, directly blocking deployment on specific hardware or systems.  
\textit{Build and Compilation Failures (C.1.i, C.1.ii)} involve missing CUDA extensions, compiler errors, or misconfigured build toolchains. For example, PyTorch fails to detect CUDA after a CPU-only installation because of unresolved GPU dependencies~\cite{c1case1}.  
\textit{Dependency Errors (C.1.iii, C.1.iv)} stem from ABI incompatibilities, outdated packages, or broken CI pipelines, often causing inconsistent builds across environments. One user resolved missing GPU libraries by manually aligning CUDA and cuDNN versions with framework requirements~\cite{c1case2}.  
\textit{Hardware Adaptation Bugs (C.1.v, C.1.vi)} include device recognition and compatibility failures when deploying to different hardware backends. For instance, DeepSpeed crashes due to a mismatch between the quantizer configuration and the underlying accelerator~\cite{c1case3}.  
These failures typically occur early in the development lifecycle and reflect the fragility of framework-environment integration, especially under diverse hardware and deployment conditions.\looseness=-1

\textbf{\faBug\ 2.Resource Efficiency and Memory Management (C.2).}
\label{taxonomy:c.2}
This theme includes 2,656 reports (10.33\% of all bug-related issues) reflecting inefficiencies in computation, memory usage, and precision control during LLM training and inference. These issues cause improper memory release, frequent OOMs, memory leaks, or abnormal resource scheduling, severely degrading training performance.  
\textit{Scheduling and Parameter Management Errors (C.2.i)} involve coordination failures in setups such as pipeline parallelism, ZeRO-offload, and long-sequence training, leading to broken gradient paths or missing parameter states. For example, a user reports a \texttt{NoneType.backward} error during large-batch training with Transformers and DeepSpeed~\cite{c2case1}.  
\textit{Performance Degradation under Resource Scaling (C.2.ii)} covers cases where training slows despite increased hardware resources, as when a DeepSpeed user observes reduced throughput after increasing batch size on LLaMA-7B due to suboptimal scheduling or contention~\cite{c2case2}.  
\textit{Quantization and Precision Impact (C.2.iii)} includes inference failures from precision-induced errors, such as enabling BF16 with 4-bit quantization causing repetitive token generation and reduced output diversity~\cite{c2case3}.  
\textit{Memory Fragmentation and Leakage (C.2.iv)} refers to unreleased memory blocks, cumulative GPU memory growth, and fragmentation over long runs, often leading to OOM crashes or unstable performance.  
Collectively, these sub-themes highlight the sensitivity of LLM frameworks to scheduling misalignments, numerical instability, and memory inefficiencies, especially in large-scale, high-throughput training.\looseness=-1

\begin{figure*}[htbp]
     \centering
    \includegraphics[width=0.8\textwidth]{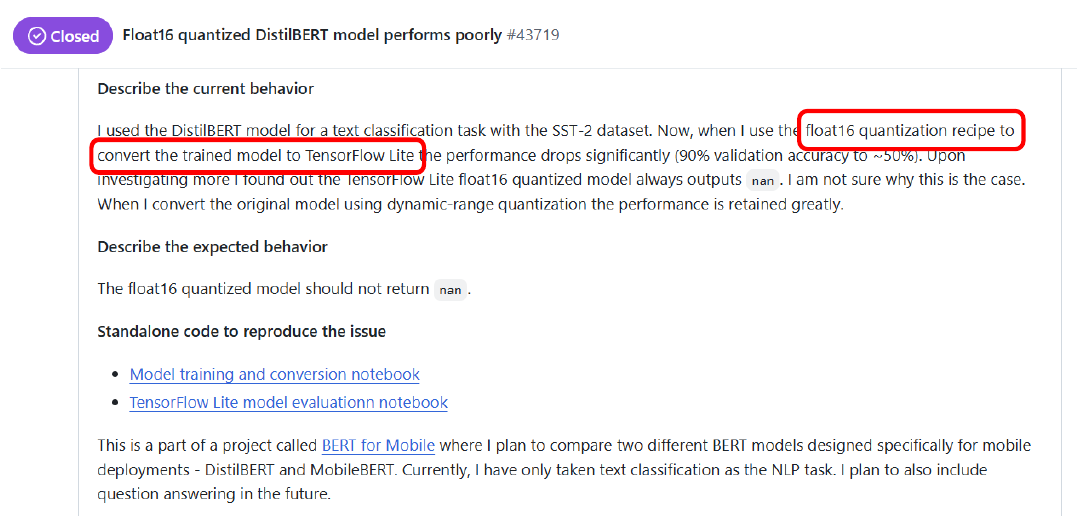}
    \vspace{-4mm}
     \caption{A sample post in Precision-Induced Accuracy Degradation (C.4.ii)}   
     \vspace{-4mm}
     \label{fig:C42}
\end{figure*}

\begin{figure*}[htbp]
     \centering
    \includegraphics[width=0.8\textwidth]{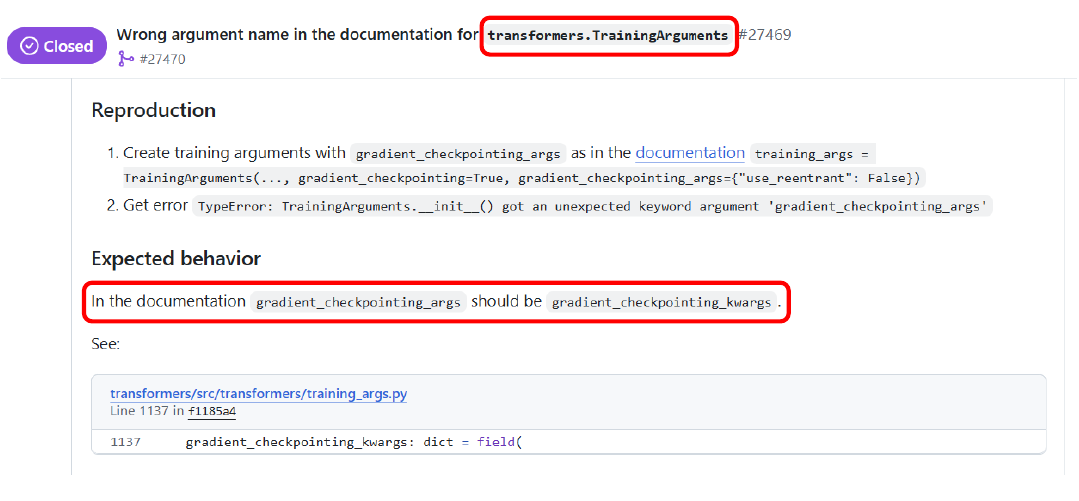}
    \vspace{-4mm}
     \caption{A sample post in Annotation and Metadata Inconsistencies (C.10.ii)}   
     \vspace{-4mm}
     \label{fig:C102}
\end{figure*}

\textbf{\faBug\ 3.Distributed and Parallel Execution (C.3).}
\label{taxonomy:c.3}
This theme includes 2,133 reports (8.30\% of all bug-related issues) arising from failures in communication protocols, synchronization logic, and multi-level parallelism during distributed training. These issues manifest as communication timeouts, node desynchronization, or process hangs, leading to inconsistent outcomes or task failure.  
\textit{Communication and Initialization Failures (C.3.i)} involve misconfigurations that block process group formation, such as incorrect \texttt{world\_size}, backend mismatches, or missing environment variables. For example, a PyTorch user reports NCCL errors in version 2.1.0 that disappear after downgrading to 2.0.1~\cite{pytorch_nccl_issue}.  
\textit{Collective Operation Failures (C.3.ii)} cover breakdowns in gradient communication after initialization, including hanging collectives or incorrect summation in \texttt{all\_reduce}. A TensorFlow user observes corrupted training due to inaccurate gradient sums from \texttt{nccl\_ops.all\_sum}~\cite{tensorflow_all_sum_issue}. 
\textit{Task Coordination Errors (C.3.iii)} include mismatched execution order, incomplete checkpoint states, and synchronization deadlocks. One PyTorch user reports backward pass hangs on 8 GPUs due to NCCL AllReduce coordination failure during large BERT training~\cite{c3case3}.  
\textit{Hybrid Parallelism Misconfiguration (C.3.iv)} refers to runtime crashes and deadlocks from invalid combinations of tensor parallelism, pipeline parallelism, and optimization strategies such as ZeRO or FSDP.  
Collectively, these sub-theme underscore the fragility of distributed training infrastructures, where minor configuration errors or misaligned execution states can cascade into hard-to-debug failures that stall or corrupt large-scale runs.\looseness=-1

\textbf{\faBug\ 4.Numerical Stability and Precision Management (C.4).}
\label{taxonomy:c.4}
This theme includes 1,530 reports (5.95\% of all bug-related issues) involving unstable numerical operations, low-bit computation errors, and precision mismatches across training and inference. These issues manifest as precision loss, overflows, NaNs, or inconsistent outputs, degrading model performance or causing crashes.  
\textit{Numerical Instability and Operator Errors (C.4.i)} involve NaNs, Infs, or gradient explosions from unstable operations such as division by near-zero values or unchecked exponentials. For example, a PyTorch user reports NaN losses after a few epochs when using \texttt{autocast} and \texttt{GradScaler} in mixed-precision training~\cite{c4case1}.  
\textit{Precision-Induced Accuracy Degradation (C.4.ii)} covers accuracy drops or regressions caused by low-precision execution, including truncation errors, type mismatches, or underflow. A TensorFlow user finds that a float16-quantized DistilBERT model suffers major accuracy loss, revealing the fragility of quantized inference workflows~\cite{tensorflow_fp16_issue}.  
\textit{Checkpoint Precision Compatibility (C.4.iii)} refers to loading failures from mismatched precision between saved and target environments. For instance, a Diffusers user cannot resume training from an FP16 LoRA DreamBooth checkpoint, encountering an ``Attempting to unscale FP16 gradients'' error~\cite{c4case3}.  
These bugs often emerge silently and are hard to diagnose from surface symptoms, yet they critically affect training convergence, reproducibility, and checkpoint portability in LLM workflows.\looseness=-1

\textbf{\faBug\ 5.Training and Inference Control Logic (C.5).}
\label{taxonomy:c.5}
This theme includes 4,314 reports (16.78\% of all bug-related issues) involving flaws in training strategy execution, model state management, and inference behavior control. These issues cause abnormal behaviors such as non-updating loss, gradient errors, failed initialization, or incorrect outputs, directly affecting model quality and usability.  
\textit{Training Strategy and Optimizer Usage Errors (C.5.i)} cover misconfigured warm-up schedules, faulty parameter group logic, or inconsistent optimization behavior. A PyTorch user finds that FSDP yields different gradient norms than DDP on T5-Large, causing norm clipping to diverge training outcomes~\cite{c5case1}.  
\textit{Training Stability Failures (C.5.ii)} involve unstable loss curves or divergence during fine-tuning. For example, a user reports loss instability when training Mistral-7B using HuggingFace Transformers with the HF Trainer, not observed in direct scripts~\cite{c5case2}.  
\textit{Initialization and State Configuration Errors (C.5.iii)} include uninitialized weights, missing state toggles, and faulty training-evaluation transitions. A PyTorch user reports that FSDP fails to initialize on the meta device, breaking memory-efficient LLM loading~\cite{c5case3}.  
\textit{Gradient and Optimizer Control Errors (C.5.iv)} capture missing gradient clipping, incorrect accumulation steps, or inconsistent learning rate updates. A DeepSpeed user attributes loss fluctuations to faulty gradient accumulation logic~\cite{deepspeed_gradient_accumulation}.  
\textit{Inference Control Failures (C.5.v)} refer to faulty decoding behavior such as constant or meaningless outputs during or after training. A Megatron user observes nonsensical generations mid-training, suggesting inference state corruption~\cite{megatron_inference_output}.  
These bugs often emerge late in training, are hard to isolate, and severely compromise convergence stability, reproducibility, and the trustworthiness of LLM inference results.\looseness=-1

\textbf{\faBug\ 6.Data Pipeline and I/O Management (C.6).}
\label{taxonomy:c.6}
This theme includes 1,030 reports (4.01\% of all bug-related issues) involving failures in data loading, preprocessing, checkpointing, and I/O consistency across LLM workflows. These issues cause deadlocks, format or path errors, cache failures, and corrupted checkpoints, disrupting training or inference and impacting long-text or multimodal tasks.  
\textit{Data Flow Processing Failures (C.6.i)} cover deadlocks, shuffle misbehavior, and modality misalignment in multi-worker or multimodal pipelines. For example, a PyTorch user reports a DataLoader deadlock when using multiple workers~\cite{pytorch_dataloader_deadlock}.  
\textit{Input Reading and Data Loading Errors (C.6.ii)} include invalid file paths, shape mismatches, or incomplete file reads. A LLaMA-2 user encounters connection failures when loading large raw text datasets during LoRA fine-tuning~\cite{c6case2}.  
\textit{Preprocessing and Type Handling Bugs (C.6.iii)} involve incorrect augmentations or datatype mismatches that cause runtime errors, such as \texttt{dtype} conflicts from Flash Attention 2~\cite{c6case3}.  
\textit{Checkpoint Export and Persistence Failures (C.6.iv)} capture incomplete serialization, missing submodules, or corrupted outputs during model saving. A DeepSpeed user reports missing parameters in checkpoints under ZeRO-Stage 3~\cite{deepspeed_checkpoint_issue2}.  
\textit{Cross-Version Checkpoint Compatibility (C.6.v)} includes loading failures after framework or format upgrades, such as shard mismatch errors when restoring Megatron checkpoints from earlier versions~\cite{megatron_checkpoint_bug}.  
These bugs often surface late in workflows and are difficult to detect or reproduce, yet they critically undermine reproducibility, recovery, and deployment stability in large-scale LLM training.\looseness=-1

\textbf{\faBug\ 7.Interface Design and Configuration Management (C.7).}
\label{taxonomy:c.7}
This theme includes 1,664 reports (6.47\% of all bug-related issues) arising from mismatches between framework APIs, configuration layers, and developer expectations. These issues lead to ineffective configurations, unexpected behaviors, or crashes, often appearing even when users follow documentation, thereby reducing usability and developer confidence.  
\textit{API Parameter Misuse (C.7.i)} covers invalid argument values, type mismatches, or silent misbehavior from incorrect parameter usage. For example, setting \texttt{attn\_implementation=``eager''} on LLAVA-7B yields incorrect output IDs and runtime errors~\cite{c7case1}.  
\textit{Interface Specification and Compatibility Violations (C.7.ii)} include cases where actual behavior contradicts API contracts, such as fine-tuning LLaMA 3.2 1B with Trainer and FSDP failing due to checkpoint–parameter size mismatches~\cite{c7case2}.  
\textit{Toolchain and Configuration Automation Failures (C.7.iii)} involve faulty command-line tools or pipeline integrations, such as \texttt{$is\_zero\_init\_model$} always returning \texttt{False} under ZeRO-3~\cite{c7case3}.  
\textit{Parameter Handling and Override Conflicts (C.7.iv)} refer to parsing errors or unintended overrides that disrupt runtime behavior, e.g., memory estimation mismatches from misresolved DeepSpeed configuration entries~\cite{deepspeed_memory_allocation}.  
\textit{Documentation and Logging Inconsistencies (C.7.v)} capture mismatches between guides, logs, and scripts, as seen in DeBERTa where documented and actual outputs diverge~\cite{c7case5}.  
These bugs frequently manifest as confusing runtime behavior, silent failures, or misleading diagnostics, complicating debugging and undermining reproducibility in complex LLM workflows.\looseness=-1

\textbf{\faBug\ 8.Architecture and Module Integration (C.8).}
\label{taxonomy:c.8}
This theme includes 294 reports (1.14\% of all bug-related issues) originating from defects in architecture definitions, submodule integration, and execution logic that affect both training and inference stability. These issues arise from structural incompatibility, weight loading errors, or inconsistent component behavior, preventing models from being correctly built or executed.  
\textit{Module Integration Failures (C.8.i)} include errors during sanitizer tests, static tracing, or runtime validation, such as unregistered layers, missing hooks, or backend misalignment. One PyTorch user encounters replication padding errors during sanitizer validation~\cite{pytorch_replicationpad2d_error2}.  
\textit{Mechanism Disruptions (C.8.ii)} involve breakdowns in autograd, caching, or memory sharing, often causing incorrect gradient propagation or silent state overwrites~\cite{c8case2}.  
\textit{Execution Failures (C.8.iii)} refer to crashes in forward or backward passes due to broken submodules or improperly linked components, as in a Megatron inference failure caused by uninitialized modules~\cite{megatron_example_bug}.  
\textit{Architectural Logic Errors (C.8.iv)} include shape mismatches, dimension misalignment, and layer misconfigurations that cause tensor incompatibility or distorted outputs.  
\textit{Generation Anomalies (C.8.v)} capture silent inference errors where models run without crashing but produce degenerate or repeated outputs, often signaling upstream architectural corruption~\cite{c8case5}.  
Although less frequent, these bugs undermine model integrity and are difficult to trace, as they may silently propagate through the architecture or degrade output quality without explicit errors.\looseness=-1

\textbf{\faBug\ 9.Code Quality, Logging, Testing, and Security (C.9).}
\label{taxonomy:c.9}
This theme includes 5,071 reports (19.72\% of all bug-related issues) exposing deficiencies in code correctness, test robustness, debugging support, and runtime stability. These issues range from code typos and missing test coverage to misleading logs, persistent CI failures, and exposed security risks, undermining maintainability and increasing the likelihood of undetected high-risk faults.  
\textit{Testing and Quality Assurance Gaps (C.9.i)} include missing test coverage, unstable CI pipelines, or insufficient sanitizer validation. A PyTorch user reports a \texttt{ReplicationPad2D} configuration error under compute sanitizers, revealing test blind spots~\cite{c9case1}.  
\textit{Code and Algorithm Logic Failures (C.9.ii)} involve broken control flows, type mismatches, or silent deviations. A Megatron user encounters a \texttt{NoneType} shape error during training due to flawed computation graph logic~\cite{c9case2}.  
\textit{Boundary Condition and Error Handling Issues (C.9.iii)} refer to missing checks for zero-sized batches or undefined state transitions. One user reports that \texttt{torch.nn} modules fail with zero-batch inputs~\cite{c9case3}.  
\textit{Logging and Debugging Failures (C.9.iv)} capture discrepancies between reported and actual behavior caused by missing instrumentation or misleading logs. A DeepSpeed user observes large gaps between reported and actual memory usage~\cite{c9case4}.  
\textit{Runtime Crashes and Device-Level Errors (C.9.v)} include kernel failures, segmentation faults, or device faults that halt execution, as seen in a Megatron multi-GPU training crash~\cite{c9case5}.  
\textit{Security and Stability Bugs (C.9.vi)} involve nondeterministic behavior or unsafe state handling, especially after interruptions or checkpoint resumes. A Megatron user observes inconsistent loss values after resuming MoE training~\cite{c9case6}.  
These bugs often stem from inadequate testing, limited observability, or unsafe fallback logic. While not always tied to core algorithms, they significantly weaken framework robustness, debuggability, and production readiness.\looseness=-1

\textbf{\faBug\ 10.Documentation and Maintainability (C.10).}
\label{taxonomy:c.10}
This theme includes 749 reports (2.91\% of all bug-related issues) highlighting deficiencies in documentation quality, metadata accuracy, toolchain integration, and instructional reliability. These issues hinder learning, debugging, and long-term maintenance, often forcing users into costly trial-and-error workflows.  
\textit{Documentation Gaps and Standard Violations (C.10.i)} include missing parameter descriptions, outdated examples, or unstated platform-specific assumptions. For instance, a PyTorch user notes that the \texttt{inplace} parameter is undocumented in \texttt{torch.nn.functional.hardsigmoid}~\cite{c10case1}.  
\textit{Annotation and Metadata Inconsistencies (C.10.ii)} involve stale docstrings, misleading comments, or mismatched type hints. A Transformers user encounters a \texttt{TypeError} due to an incorrect argument name documented in \texttt{TrainingArguments}~\cite{c10case2}.  
\textit{Toolchain Synchronization Failures (C.10.iii)} cover broken documentation builds, outdated config schemas, or missing compatibility notes. One DeepSpeed user reports a \texttt{torch.compiler} import error in version 0.15.3 not reflected in the official documentation~\cite{c10case3}.  
\textit{Instructional and Example Errors (C.10.iv)} capture broken tutorials, incorrect code snippets, or misleading usage guidance. A Megatron user encounters a circular dependency in the DreamBooth tutorial between \texttt{transformer\_engine} and \texttt{core.utils}~\cite{c10case4}.  
These documentation-related failures undermine user trust, slow onboarding, and increase maintenance costs, especially when guides diverge from evolving APIs or system assumptions.\looseness=-1

\textbf{Summary.} Bug reports constitute 66.36\% of our dataset, revealing multi-layer fragility across the LLM software stack.
Environment-level compatibility failures (18.21\%, C.1) dominate early-stage breakdowns, typically triggered by mismatched dependencies, unsupported hardware backends, or failed builds during installation and initialization.
Runtime instability (16.28\%, C.2, C.4) stems from memory fragmentation, degraded throughput, numerical divergence, and quantization-induced errors, often emerging during large-batch or long-sequence training.
Distributed orchestration errors (8.30\%, C.3) arise from brittle coordination across tensor, pipeline, and FSDP components, with minor misconfigurations cascading into deadlocks, desynchronization, or corrupted states.
Training control and architectural integration bugs (17.92\%, C.5, C.8) disrupt convergence and inference behavior, manifesting as frozen loss curves, incorrect optimizer updates, uninitialized modules, or degenerate generations.
Finally, observability and maintainability weaknesses (39.29\%, C.6–C.10) dominate at scale, spanning faulty I/O pipelines, incomplete checkpoints, misleading logs, insufficient test coverage, and outdated documentation. These failures frequently surface during recovery, multi-modal data handling, or cross-version migration, indicating that even when core algorithms function correctly, peripheral system tooling remains a primary barrier to robustness.\looseness=-1

\finding{1}{
 \textbf{Finding 3.} 
    Bug reports highlight fundamental limitations of current LLM frameworks, concentrated in system compatibility, distributed orchestration, training control, and observability. Compatibility failures (18.21\%, C.1) remain the most frequent, while runtime instability in memory and precision handling (16.28\%, C.2, C.4) represents a major barrier to stable execution. Notably, observability and maintainability weaknesses (39.29\%, C.6–C.10) dominate at scale, indicating that fragility in system tooling, rather than algorithmic components, is the leading cause of large-scale failure.\looseness=-1

}

\begin{figure*}[htbp]
     \centering
    \includegraphics[width=0.95\textwidth]{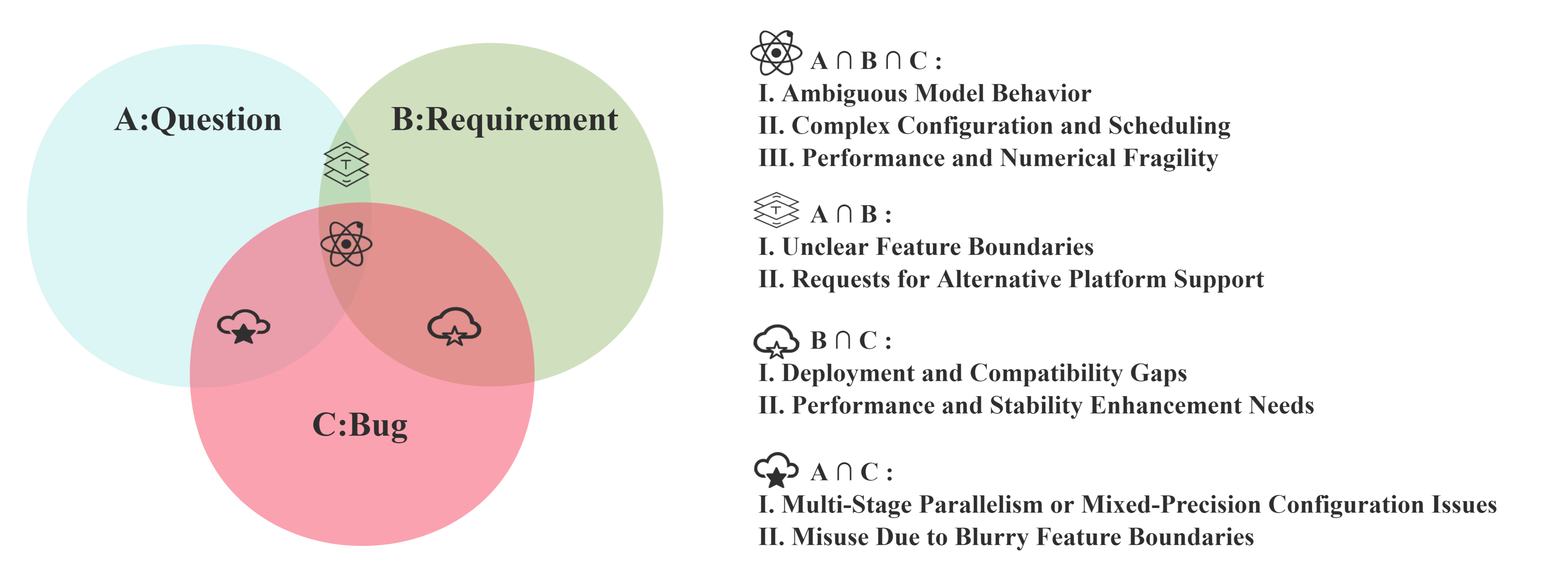}
    \vspace{-4mm}
     \caption{LLM-Centric Challenges Summarized from our Taxonomy}   
     \vspace{-4mm}
     \label{fig:Three_category_Venn_diagram}
\end{figure*}

\vspace{-2mm}
\subsubsection{Answer to RQ1.\\}

Our taxonomy organizes LLM-centric challenges in DL frameworks into three complementary dimensions: operational confusion, unmet expectations, and system-level failures. Across these categories, five recurring challenge themes emerge: ambiguous model behavior, complex configuration and scheduling, unclear feature boundaries, deployment friction in heterogeneous environments, and performance or numerical fragility. These themes cut across user roles and report types, reflecting structural limitations in how frameworks support large-scale model development. For instance, ambiguous training behaviors, such as diverging loss or repetitive outputs, appear in both user questions (A.3) and bug reports (C.5), revealing gaps in behavioral baselines and runtime observability. Complex configuration interactions across ZeRO, FSDP, and precision settings show up as confusion posts (A.2), automation requests (B.5), and silent-failure bugs (C.2). Users also struggle to determine whether certain features are supported at all, reflecting blurry module boundaries and under-documented APIs (A.2, B.2). Deployment challenges, such as missing Apple Silicon or ROCm support, manifest as compatibility bugs (C.1) or platform extension requests (B.1). Finally, difficulties in precision management and memory scheduling affect both stability and performance expectations, surfacing in performance bugs (C.2, C.4) and tuning-related requests (B.5). Together, these patterns reveal systemic bottlenecks in usability, extensibility, and robustness that must be addressed for real-world LLM development. Detailed differences between questions, requirements, and bugs are shown in Fig.~\ref{fig:Three-categoryVenndiagram}.\looseness=-1

While these challenges recur across the taxonomy, their expression in Questions, Requirements, and Bugs remains distinct, validating the need to separate conceptual confusion, unmet needs, and concrete failures. For ambiguous model behavior, Questions (A.3) reflect user uncertainty in diagnosing loss divergence or abnormal outputs, Bugs (C.5) identify training control logic violations such as faulty gradient accumulation, and Requirements (B.5) call for better convergence monitoring or debugging tools. In complex configuration and scheduling, Questions (A.2) arise when users cannot determine correct multi-stage parallelism or mixed-precision settings, Bugs (C.2, C.5) document runtime crashes or silent failures from misconfiguration, and Requirements (B.5) request auto-tuning or performance-aware defaults. For unclear feature boundaries, Questions (A.2.ii) stem from uncertainty about module applicability, while Requirements (B.2) seek more modular, extensible APIs. Deployment friction surfaces as Bugs (C.1) when existing platform builds fail and as Requirements (B.1) when users request support for new environments. Finally, performance and numerical fragility appears in Bugs (C.2, C.4) as instability after enabling quantization or large batch sizes, and in Requirements (B.5) as demands for better tuning and profiling support. Even when technical symptoms overlap, user framing, expectations, and intent vary substantially, underscoring the value of a taxonomy that captures both the problem space and the user’s perspective.\looseness=-1

\vspace{-2mm}
\section{Interview with Users and Developers}
\label{sec:interview}
Building on the taxonomy of LLM-centric questions, requirements, and bugs from Stage I, we further conduct interviews to deepen our understanding of real-world challenges. We invite 11 LLM users who use DL frameworks to fine-tune, customize, or deploy LLMs, and eight DL framework developers responsible for building, maintaining, or testing LLM-centric functionalities. The following sections introduce how we design the interview guideline, the interviewee's background, the interview process, and the analysis of the collected insights.\looseness=-1

\textbf{Interview Guideline Design.}
To address the limitation of our taxonomy, i.e., it fails to capture subjective experiences and the rationale behind real development practice, we further design an interview guideline and collect feedback from real-world LLM users and DL framework developers. We adopt a format to balance clarity with flexibility: the interviewer follows a consistent logical flow while adapting dynamically to participants’ responses, which is essential for uncovering subtle insights, maintaining engagement, and clarifying potential misunderstandings in real time.

The guideline is organized into four clearly defined parts, each targeting a specific dimension of the taxonomy to ensure comprehensive coverage. Throughout its development, we consulted domain experts to refine the structure and verify its technical relevance. This collaborative and evidence-driven process enhances the credibility of the findings and ensures the interviews yield actionable feedback. 

The first part introduces the study’s background and objectives to align participants, while also collecting information about their experience in LLM development or deployment. The remaining sections focus on validating and expanding the taxonomy by examining whether its issue types, themes, sub-themes, and descriptions reflect the participants’ actual experiences. Open-ended questions invite participants to share new challenges and propose improvements, ensuring the taxonomy continues to capture emerging trends and overlooked issues. By mapping interview questions to specific taxonomy elements, we can assess whether these categories faithfully represent real-world challenges and reveal perspectives not visible in public issue repositories. \looseness=-1

\vspace{-2mm}
\subsection{Background of Interviewees}
We separate interviewees into two distinct groups: LLM users and DL framework developers, to capture complementary perspectives on the same set of challenges. LLM users provide experience-driven insights into practical usage scenarios, common pain points, and unmet needs, revealing how frameworks are perceived and applied in the field. In contrast, DL framework developers offer system-level knowledge of design rationale, implementation trade-offs, and internal constraints, explaining why certain issues arise or persist. This dual-role design enables cross-validation between user-reported problems and developer explanations, reduces bias toward either the usage side or the implementation side, and ensures a balanced understanding of LLM-related challenges. Such a structure not only strengthens the credibility of our taxonomy validation but also increases the likelihood of deriving actionable, technically feasible improvement strategies.

LLM users are practitioners who primarily build, fine-tune, deploy, or maintain LLMs using existing DL frameworks (e.g., PyTorch, MindSpore). Although many of them possess development skills, they do not directly contribute to the implementation or maintenance of the DL frameworks themselves. Their typical tasks include applying LLMs in specific fields like text generation or developing LLMs like prompt tuning and model inference. This group includes two professional engineers with over three years of experience in LLM-based systems, and nine computer science students with an average of two years' experience in fine-tuning open-source LLMs for research and competitions.

DL framework developers are contributors or maintainers of DL frameworks or LLM-supporting toolkits. They are responsible for the design, implementation, optimization, and bug-fixing of core components such as operator kernels, parallelism strategies, training utilities, and model definition abstractions. Their work directly impacts how LLMs can be supported at the system level. This group includes eight LLM users from leading Internet companies: two senior managers with over eight years in framework validation, five engineers focused on LLM-centric feature development (three for training, two for inference), and one developer working on LLM deployment optimization. Additional information is available on our website~\cite{sharelink}. \looseness=-1

\vspace{-4mm}
\subsection{Interview Process} 
The first author conducts online interviews with 11 LLM users and eight DL framework developers, following the designed guidelines. Each 30–40 minute session combines background briefings with ranking, short-answer, and open-ended questions to validate and extend the taxonomy, organized into four parts directly mapped to its core dimensions. This structure ensures comprehensive coverage while keeping responses comparable across participants, and the interview format allows flexible follow-ups to capture subtle, context-specific challenges often missing from public reports. Interviewing both LLM users and framework developers brings complementary perspectives, addressing not only symptom-level problems but also underlying causes and feasibility constraints, which strengthens the link between user needs and framework capabilities. Open coding and collaborative review further ensure emergent insights are systematically captured, categorized, and cross-validated, enhancing credibility, minimizing bias, and supporting reproducibility.\looseness=-1

\textbf{\underline{Part I.}}
We begin by introducing the study's background and goal, as well as understanding the challenges in supporting LLMs of DL frameworks. We then collect demographic information, including participants' roles, educational background, LLM experience, and the frameworks they use or develop. We also ask about their specialization in LLM types (e.g., text or multimodal) and tasks such as inference, quantization, training, or fine-tuning.\looseness=-1

\textbf{\underline{Part II.}}
This section examines common issues users encounter when adopting DL frameworks for LLMs and the perspectives of developers on these questions. Users rank the question types identified in Stage I, describe other frequent questions, and share overall experiences. Developers assess the importance of user questions and share the challenges they encounter when handling user-reported issues. \looseness=-1

\textbf{\underline{Part III.}}
We focus on framework bugs that affect LLM quality and their fixing priorities. Users rank typical bug types, report missing ones, and describe memorable debugging experiences. Developers evaluate bug-fixing priorities, add unlisted bugs, and share quality assurance challenges.\looseness=-1

\textbf{\underline{Part IV.}}
We examine expectations for improving LLM support. Users rank requirements, suggest missing features, and describe unmet needs. Developers share current or planned improvements and reflect on long-term directions, plans, and goals for enhancing the framework in LLM scenarios.\looseness=-1

At the end of each session, we thank participants and outline the next steps of the study.\looseness=-1

\textbf{Transcription and Open Coding.}
After the interviews, the first author transcribes the recordings and performs open coding using NVivo~\cite{nvivo2024}. The coding process captures both predefined themes and emergent insights across four dimensions: (1) participant background and framework usage; (2) common issues and core functionalities; (3) critical bugs and debugging experiences; and (4) expectations and improvement suggestions. The second author reviews and refines the code. Both authors independently group the codes into thematic categories and then jointly finalize the results, ensuring consistency and minimizing bias.\looseness=-1

\begin{figure*}[htbp]
     \centering
    \includegraphics[width=1.02\textwidth]{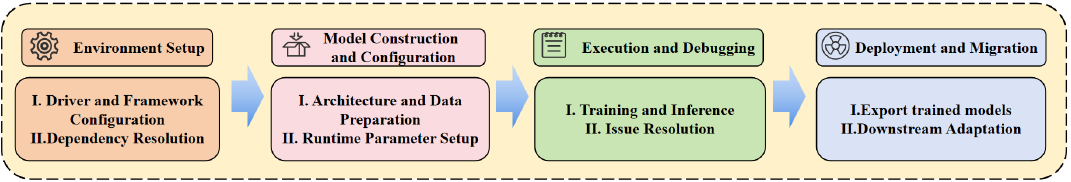}
    \vspace{-6mm}
     \caption{LLM development lifecycle}   
     \vspace{-6mm}
     \label{fig:LLM_development_lifecycle}
\end{figure*}

\subsection{Result Analysis: Interviews with Users and Developers}

\subsubsection{Interview Results from Users\\}

We present interview findings from LLM users, focusing on their perspectives on the requirement, question, and bug categories in our taxonomy. Regardless of seniority, all interviewees agree that LLM development typically follows four high-level stages: (1) \textit{environment setup} for driver installation, framework configuration, and dependency resolution; (2) \textit{model construction and configuration} for preparing model structures, datasets, and runtime parameters; (3) \textit{execution and debugging} for training, inference, and resolving functional or performance issues; and (4) \textit{deployment and migration} for exporting trained models and adapting them to downstream tasks, as shown in Fig.~\ref{fig:LLM_development_lifecycle}. They unanimously confirm that our taxonomy’s themes and sub-themes are representative and align with real workflows, with different challenge types arising at different stages.\looseness=-1

\textbf{Question Type.}
All interviewees affirm the relevance of our question taxonomy, agreeing that its five themes capture common sources of confusion. A.1 (installation issues) and A.4 (deployment errors) occur mainly in \textbf{Stages 1} and \textbf{4}, while A.3 (unexpected behavior) and A.5 (strategy confusion) dominate \textbf{Stage 3}. \looseness=-1

One user finds the boundary between \textit{System Support and Dependency Management} (A.1.i) and \textit{Configuration and Parameter Usage} (A.1.ii) unclear. They describe installing PyTorch via \texttt{pip} but encountering GPU failures due to a missing \texttt{libcuda.so}, suggesting renaming A.1.i to \textit{System Environment and Dependency Setup} and A.1.ii to \textit{Runtime Configuration and Launch Parameters} to better distinguish pre- and post-installation failures.\looseness=-1

Another interviewee questions the clarity between \textit{Usage Gaps} (A.2) and \textit{Unexpected Behavior} (A.3). They suggest A.2 cover confusion before successful execution (e.g., enabling mixed precision) and A.3 address post-execution uncertainty (e.g., abnormal loss despite correct setup). On A.3.iii \textit{Model Behavior and Output Verification}, they stress the need for diagnostic support and behavioral baselines, citing MoE training with Megatron where suspicious outputs appear without errors.\looseness=-1

Interviewees frequently report difficulty configuring hybrid strategies such as mixed parallelism, parameter freezing, or optimizer–scheduler combinations, issues not caused by bugs or missing docs, but by absent conceptual guidance. This motivated A.5 \textit{Training Strategy and Configuration Confusion}. A.5.i covers misunderstandings when combining advanced techniques (e.g., LoRA with ZeRO, mixed precision with freezing), where symptoms like stagnant loss or ineffective gradients appear without clear cause. A.5.ii covers regressions or instability after framework upgrades due to undocumented defaults or backward incompatibilities. Many rank A.3 and A.5 as the most time-consuming, noting that without diagnostics, root causes of repeated outputs, gradient anomalies, or memory crashes remain unclear. Frameworks rarely document optimizer–scheduler–parallelism–precision interactions, leading to silent errors that degrade performance after extensive experimentation.\looseness=-1

Installation issues (A.1) are early hurdles, declining with experience. Newcomers, especially with under one year of exposure, consistently rank A.1 highest in frustration, citing obscure errors and vague guidance. One recalls struggling to install MindSpore’s MindSpeed on Ascend for Yi-6B debugging, where missing Apex support and version conflicts consumed hours. Experienced users raise fewer setup concerns but point to undocumented assumptions, such as \texttt{torch.save} with DeepSpeed ZeRO3 silently saving only partial weights.\looseness=-1

Platform deployment issues (A.4) disrupt workflows when switching devices or integrating inference toolchains, but are seen as less critical than runtime confusion. Operational questions (A.2) are often resolved via community examples. Across all experience levels, usage gaps, caused by unclear documentation, fragile configuration logic, or implicit behaviors, remain a dominant barrier, demanding guidance that is up-to-date, context-aware, and workflow-aligned.\looseness=-1

One user fine-tuning a 6B-parameter model with MindSpore’s MindSpeed described severe friction debugging divergence. The tutorial resembled internal engineering notes, referencing undocumented environment variables, omitting runnable examples, and assuming parallel training expertise. Critical behaviors, such as \texttt{ms.save\_checkpoint} silently saving only a single shard under \texttt{slice\_mode=auto\_parallel}, were undocumented, causing days of confusion and misdiagnosed failures. Without standard logging or visual debuggers, they modified framework source to trace tensors:\looseness=-1 

``\textit{The so-called debugging guide felt like a private memo between core developers. If you’re not already in the loop, you’re completely lost.}''\looseness=-1

\textbf{Requirement Type.}
Participants consistently affirm that our requirement taxonomy captures key challenges in LLM development. One interviewee observes that the original B.2, \textit{Functionality Support for LLM Development}, mixes high-level system capabilities (e.g., API design, distributed training, model compatibility) with low-level efficiency needs (e.g., performance tuning, workflow configuration), complicating interpretation and maintenance. To address this, we split B.2 into \textit{System and Model Feature Support} (modular APIs, distributed execution, data ingestion, model format compatibility) and B.5 \textit{Training Workflow and Efficiency Optimization} (runtime tuning, memory usage, mixed precision, long-sequence support), as detailed in Section~\ref{taxonomy:b.2} and Section~\ref{taxonomy:b.5}. Lifecycle-wise, B.1 (environment compatibility) and B.4 (documentation support) are most relevant in \textbf{early stages}, while B.2 and B.5 dominate \textbf{Stage 3}. B.3 (code reliability) is seen as foundational and spans multiple stages.\looseness=-1

Interviewees propose two new sub-themes. First, beyond hardware and dependency compatibility in B.1, developers face growing challenges in cross-framework and cross-tool interoperability, such as converting models between HuggingFace and Megatron, exporting to ONNX or TensorRT, and adapting configurations across DeepSpeed and vLLM. We add B.1.iv, \textit{Multi-platform Interoperability}, to address this. Second, beginners often encounter errors due to outdated examples or deprecated configurations, prompting B.4.iii, \textit{Example Drift and Documentation Mismatch}, under the community and documentation category.\looseness=-1

Beginners and general users rank B.1 as the most urgent, citing installation failures, dependency conflicts, and debugging difficulties on CPU-only systems, Windows, or non-Docker setups, which delay onboarding and erode confidence. Experienced developers prioritize B.2 and B.5 for scaling training, integrating custom modules, and maintaining efficient workflows. They emphasize modular APIs, flexible checkpointing, and memory-efficient training. B.3, \textit{Code Reliability and Testing Assurance}, is valued for long-term maintainability, with requests for better logging, debugging tools, and fault-tolerant training. While B.4 is less critical in daily operations, many, especially beginners, stress its role in onboarding, troubleshooting, and bridging documentation gaps.\looseness=-1

One user reports silent loss stagnation after resuming Megatron-based training from a checkpoint because \texttt{load\_state\_dict} skipped mismatched optimizer weights without warning. The issue went unnoticed for days due to missing diagnostics or validation. Later, exporting the same model to ONNX failed due to undocumented custom ops in attention layers, absent in both the framework and conversion toolchain. As they summarized: \looseness=-1

``\textit{You spend weeks fine-tuning a model, only to find the checkpoint is broken and deployment fails, and the framework never tells you what went wrong.}''\looseness=-1

\textbf{Bug Type.}
Interviewees agree that our bug taxonomy captures core challenges and maps well to real-world LLM development stages. C.1 (system compatibility) and C.10 (documentation issues) dominate \textbf{Stage 1}, while C.2 (resource efficiency), C.3 (distributed execution), and C.5 (control logic) are most prevalent in \textbf{Stage 3}. They view the classification as comprehensive and relevant, yet emphasize the urgent need for stronger diagnostics, including tools to trace optimizer behavior, detect silent loss-scaling failures, and address mixed-precision instability.
Prioritization varies with experience. Novices highlight C.1, C.6 (data pipeline and I/O), and C.10 as early blockers, citing installation failures, data shape mismatches, and missing configuration details. Experienced users focus on C.2, C.3, and C.8 (architecture and module integration) to ensure training correctness, scaling stability, and deployment success. Across all levels, participants find the taxonomy effective for identifying and reasoning about framework-level bugs, while stressing that actionable diagnostics remain a critical gap.\looseness=-1

One user described installing a framework on an NVIDIA A100 without error, only for training to crash on the first forward pass. After extensive debugging, they traced it to an undocumented mismatch between the driver’s compute capability and the prebuilt CUDA binary. Later, inference produced inconsistent results until \texttt{model.eval()} was added to the export script. Restoring a LoRA fine-tuned checkpoint then failed silently, as new parameter heads were not registered during deserialization, leaving downstream tensors uninitialized. As they concluded: \looseness=-1

``\textit{You’re never told what went wrong. It’s not just one bug, it’s the chain of invisible assumptions that breaks everything.}''\looseness=-1

\finding{1}{
 \textbf{Finding 4.} 
   Interviews with LLM users show that challenges align closely with the four stages of the LLM development lifecycle: environment setup, model construction and configuration, execution and debugging, and deployment and migration. Newcomers struggle mainly in the early stages with vague installation errors, dependency mismatches, and insufficient onboarding guidance. Experienced users report Stage 3 as the most problematic, citing undocumented behaviors, silent failures, and implicit design assumptions that complicate debugging. Across all levels, participants note that documentation rarely explains how configuration parameters interact, forcing trial-and-error workflows. They call for intelligent, context-aware tooling, such as configuration validation, behavior visualization, and experiment tracking, to reduce manual effort and improve transparency. Bug priorities also vary: beginners are blocked by setup failures and data I/O issues, while advanced users face numerical instability, gradient divergence, and mixed-precision execution failures. Regardless of experience, all participants stress the need for clearer feedback, stronger runtime diagnostics, and better support for debugging, error localization, and behavioral verification throughout the lifecycle.\looseness=-1
}

\subsubsection{Interview Results from Developers\\}

We present the interview results with DL framework developers, focusing on their opinions regarding the questions, requirements, and bugs in our constructed taxonomy. \looseness=-1

\textbf{Question Type.} 
Framework developers agree that our five question categories, from installation barriers to training strategy confusion, capture the full spectrum of real-world support requests. They confirm that the taxonomy spans both early-stage issues (A.1 installation failures) and advanced challenges (A.5 training instability) and value the distinction between \textit{usage confusion} (A.2) and \textit{unexpected behavior} (A.3), which aligns with their internal triage process. In practice, most user questions are either clarification requests or suspicions of underlying bugs.\looseness=-1

Developers consistently rank \textbf{A.3 Unexpected Behavior and Unintuitive Design} as the most critical, as they involve misleading training signals, unstable gradients, diverging losses, or silent regressions, that users struggle to attribute to misconfiguration, model variance, or framework faults. The absence of expected behaviors and diagnostic support makes these cases particularly hard to reproduce and resolve, undermining trust and reproducibility. A.1 (Unclear Installation Guidance) and A.4 (Cross-Platform Deployment Issues) are also high-priority because dependency mismatches, missing packages, or incompatible hardware often block onboarding or deployment, especially in heterogeneous or resource-constrained environments. By contrast, A.2 and A.5 appear more frequently but are rated lower in severity; they typically reflect documentation and usability gaps, such as unclear parameter semantics, missing examples, or complexity in scheduling, multi-stage optimization, and distributed strategies. Several developers express interest in validation layers or AI-powered assistants to proactively check configurations and reduce user confusion.\looseness=-1

One developer described a silent failure case where ``\textit{training appears normal, but the loss remains exactly zero across all iterations.}'' The setup used DeepSpeed ZeRO-2 with automatic mixed precision (AMP) under distributed training. No errors were logged, yet overflow detection silently failed on specific devices, disabling the loss scaler without warning and suppressing gradients on those ranks. The user initially suspected model or data issues, but the root cause was a subtle AMP–gradient partitioning interaction. Without diagnostics to detect rank-level inconsistencies, the job consumed significant compute without learning. As the developer noted: ``\textit{A framework should never make training look healthy when learning isn’t actually happening.}''\looseness=-1

\textbf{Requirement Type.}
Framework developers agree that the proposed requirement taxonomy captures both the breadth and granularity of user expectations in LLM development. The five categories, from environment compatibility (B.1) to workflow optimization (B.5), address functional needs as well as non-functional concerns such as maintainability, portability, and scalability. They particularly value the clear separation between system-level enablers (e.g., B.1 hardware and deployment support) and user-facing pain points (e.g., B.2 functionality gaps and B.3 reliability issues), which aligns with their internal planning priorities.\looseness=-1

When ranking urgency, developers place B.2 and B.1 at the top. B.2 \textit{Functionality Support for LLM Development} is viewed as indispensable, as users increasingly require full-stack pipelines with flexible APIs, extensible modules, and format interoperability. Key capabilities include scalable distributed training (B.2.ii), large-scale data ingestion (B.2.iii), and checkpoint compatibility (B.2.iv). As one developer notes, ``You can’t even begin meaningful LLM training if the framework doesn’t support your parallelism plan or can’t load your checkpoint.'' B.1 \textit{New Environment Compatibility and Deployment} is equally urgent for applied deployments, with growing demand for portability across CPUs, NPUs, XPUs, and mobile devices. Missing support for ROCm, Apple Silicon, or CPU-only debugging limits adoption, while lightweight and reproducible setups (B.1.iv) are critical for onboarding and rapid iteration. B.5 \textit{Training Workflow and Efficiency Optimization} is seen as strategically important for memory efficiency, fine-grained tracing, and long-sequence training, but often follows core functionality in roadmap priority. B.3 \textit{Code Reliability in LLM Support} is recognized as foundational, yet improvements in logging, exception handling, and testing are typically deferred unless triggered by major failures. B.4 \textit{Community and Documentation Support} is valued for scaling the user base, especially through up-to-date documentation (B.4.i) and reproducible examples (B.4.iii), but progress depends heavily on sustained user feedback and contributions.\looseness=-1

One developer described the heavy engineering burden of adapting frameworks to LLM workloads on emerging hardware. Building custom operators with low-level toolchains such as Ascend C required manual memory alignment, device-specific instruction tuning, and coping with inconsistent semantics across backends. The process was time-consuming, error-prone, and poorly supported by debugging or verification tools. As the interviewee put it, ``\textit{Modern frameworks should abstract away hardware-specific quirks. If I have to hand-optimize every operator for every chip, scalability is dead on arrival.}'' This case highlights the urgent need for stronger hardware abstraction, portable operator libraries, and robust high-level APIs, core concerns captured under B.1 and B.2 in our taxonomy.\looseness=-1

\textbf{Bug Type.} 
Framework developers agree that our taxonomy of LLM-centric bugs accurately reflects real-world challenges and follows sound structural logic. They group C.1–C.6 as execution-path failures that disrupt model lifecycles during building, training, inference, or checkpointing, including system incompatibility (C.1), memory and scheduling issues (C.2), distributed failures (C.3), numerical instability (C.4), flawed training logic (C.5), and I/O disruptions (C.6). These bugs block progress and require immediate intervention. In contrast, C.7–C.10 reveal weaknesses in toolchain support and engineering maturity, configuration errors (C.7), fragile module integration (C.8), inadequate testing and logging (C.9), and incomplete or misleading documentation (C.10), which impair usability, debugging, and long-term maintainability.\looseness=-1

Developers suggest refining some category names: C.5 to ``Training Strategy and Execution Control'' to stress optimizer configuration, gradient handling, and initialization logic; C.8 to ``Framework Module Implementation and Integration'' to emphasize design and autograd consistency. They also propose two new subtypes: C.2.iv ``Memory Fragmentation and Leak,'' covering unfreed memory, fragmentation, and gradual growth leading to OOM in long runs; and C.3.iv ``Hybrid Parallelism Misconfiguration,'' for crashes caused by invalid combinations of tensor, pipeline, and optimizer parallelism.\looseness=-1

In terms of priority, C.5, C.3, and C.2 are most urgent. C.5 bugs cause silent training failures, unstable convergence, or invalid outputs, undermining correctness and requiring diagnostics to detect. C.3 and C.2 issues, deadlocks, collective mismatches, memory leaks, compromise scalability and efficiency. C.1 remains critical during onboarding when device recognition or environment setup fails. C.7 and C.10 are important for ecosystem health but are often addressed only when cascading failures occur. C.6 and C.9 are foundational yet handled reactively, while C.8 bugs are rare but highly disruptive.\looseness=-1

One developer described recurring memory scheduling bottlenecks with dynamically shaped inputs in large-scale distributed training. Prolonged stalls occurred at the dispatch stage due to repeated workspace reallocations, leaving GPUs underutilized and throughput degraded. Poor parallelism strategies and bandwidth contention amplified the problem, while BF16-based AllReduce and ReduceScatter operations introduced non-deterministic behaviors that broke reproducibility under identical configurations. In some cases, regressions stemmed from subtle operator mismatches or hardware sensitivity rather than code changes. As the developer put it, ``\textit{The dispatch stage is where things fall apart, your GPU just sits there doing nothing.}'' These failures, compounded by missing actionable runtime logs during NIC crashes or HCCL breakdowns, illustrate core challenges in C.2 (Resource Efficiency and Memory Management) and C.3 (Distributed and Parallel Execution), underscoring the need for stronger runtime observability, fault isolation, and distributed execution resilience in modern LLM frameworks.\looseness=-1

\finding{1}{
 \textbf{Finding 5.} 
    Framework developers confirm that our taxonomy captures the core challenges in enabling LLM development. For user questions, they stress the prominence of \textit{unexpected behavior and unintuitive design} (A.3), which often involves misleading training signals and silent failures that are difficult to trace without robust internal diagnostics. In requirements, they prioritize \textit{functionality support} (B.2) and \textit{new environment compatibility} (B.1), citing the need for flexible parallelism strategies, model format interoperability, and deployment across heterogeneous platforms. While they recognize the value of \textit{training workflow and efficiency optimization} (B.5), they note that it often depends on deeper architectural refactoring and therefore progresses more slowly. For bugs, developers report that advanced parallelism frequently exposes weaknesses in training control (C.5), distributed execution (C.3), and memory scheduling (C.2). They emphasize that non-crashing defects, such as hybrid parallelism misconfiguration, insufficient runtime logging, and documentation mismatches, are equally damaging, as they degrade usability and maintainability while remaining hard to detect. Across all categories, developers observe that large-scale training amplifies existing framework fragilities, reinforcing the need for stronger abstraction layers, richer observability, and cross-platform robustness to address persistent pain points.\looseness=-1

}

\begin{figure*}[htbp]
     \centering
    \includegraphics[width=1.02\textwidth]{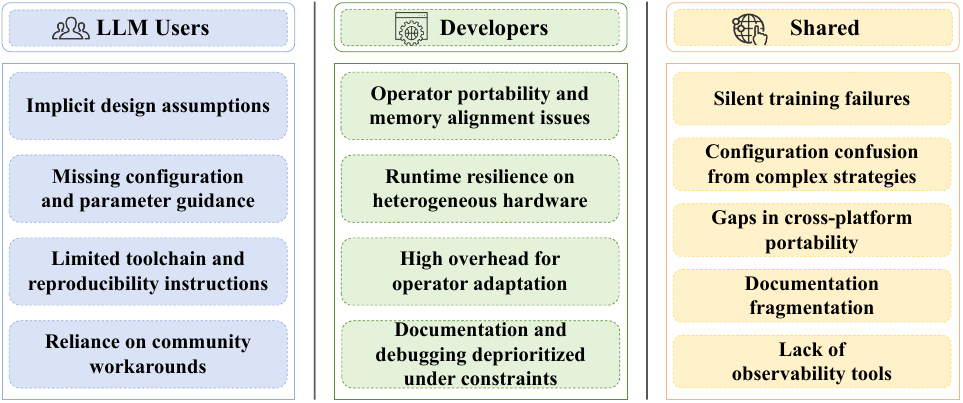}
     \caption{LLM-centric Challenges summarized from the Interview Results}   
     \label{fig:interviewresults}
\end{figure*}

\subsubsection{Answer to RQ2.\\}

\textbf{Shared and Divergent Challenges.}
Interviews with LLM users and framework developers reveal strong alignment on core limitations of current DL frameworks in supporting large-scale model development. As shown in the orange box of Fig.~\ref{fig:interviewresults}, both groups cite five recurring pain points: (1) difficulty diagnosing silent failures and unexpected behaviors, particularly under mixed precision and hybrid parallelism, where loss stagnates or gradients vanish without warnings; (2) configuration complexity when combining training strategies, with little guidance on safe interactions; (3) limited portability and weak hardware abstraction, leading to migration failures across GPUs, NPUs, and CPUs; (4) persistent documentation gaps, including outdated examples and unclear default behaviors; and (5) lack of standardized logging and diagnostics for distributed execution, fault isolation, and rank-level analysis. These issues are compounded by the absence of behavioral baselines and actionable runtime feedback, resulting in wasted compute and reduced trust in training correctness.\looseness=-1

Despite these shared concerns, focus diverges by role. Developers emphasize low-level system reliability, operator portability, memory alignment, and runtime resilience across heterogeneous backends, while users prioritize workflow transparency, usability, and configuration clarity. Developers view hardware abstraction, composable parallelism, and modular APIs as critical enablers, yet acknowledge that logging, debugging, and documentation are often deprioritized under engineering constraints, as shown in the green box of Fig.~\ref{fig:interviewresults}. LLM users, by contrast, frequently encounter undocumented behaviors, implicit design assumptions, and opaque error signals, forcing trial-and-error debugging and reliance on community workarounds. They seek behavioral baselines, configuration validation tools, and task-oriented documentation to lower entry barriers for effective LLM training and deployment, as shown in the purple box of Fig.~\ref{fig:interviewresults}.\looseness=-1

\textbf{Alignment with Taxonomy.}
Comparing annotated issue reports with interview insights confirms that our taxonomy’s three main categories, user questions, user requirements, and framework bugs, capture the most salient challenges. High-priority interview themes correspond to frequent report categories, such as A.3 (Unexpected Behavior), B.2 (Functionality Support), and C.5/C.2/C.3 (Training Control, Resource Management, Distributed Execution). However, interviews also expose underrepresented yet high-impact categories, including A.5 (Training Strategy Confusion) and C.10 (Documentation and Maintainability), which disproportionately affect novices but are rarely reported due to reproducibility difficulty. Conversely, some frequent report categories (e.g., A.1, B.4) are deprioritized by experienced users who mitigate them through tooling or shared knowledge. Requirement priorities also vary by experience: novices emphasize compatibility, setup, and documentation (B.1, B.4, C.1), while advanced users focus on scalability and robustness (B.2, B.5, C.2–C.3). Integrating both sources ensures that the taxonomy remains empirically grounded while addressing maturity-dependent and latent challenges.\looseness=-1

\section{Implications}

\begin{figure*}[htbp]
     \centering
    \includegraphics[width=1.02\textwidth]{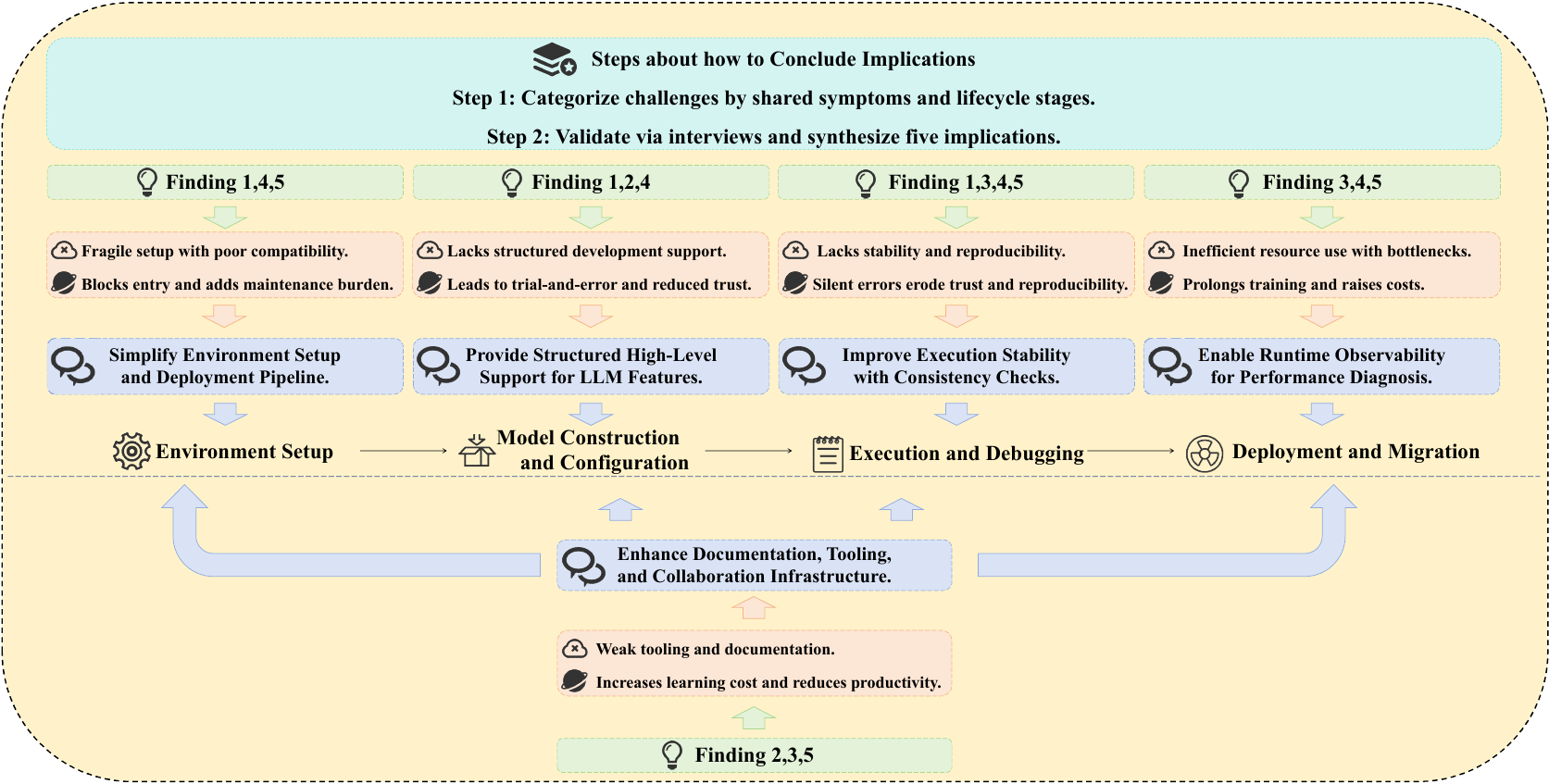}
    \vspace{-4mm}
     \caption{Implications summarized from Findings in RQ1 and RQ2}   
     \vspace{-4mm}
     \label{fig:Implications}
\end{figure*}

\subsection{Key Challenges in DL Frameworks}

Based on our taxonomy and interviews, we identify five key challenges that limit DL frameworks in supporting LLMs across different stages of the LLM development lifecycle. We begin by grouping the challenges, along with their associated themes and sub-themes in our taxonomy, according to shared symptoms and the specific lifecycle stages affected, as reflected in the findings. We then validate these grouped challenges using interview data from both LLM users and framework developers. Through this abstraction process, we derive five overarching implications, each aligned with one of the four core stages of the LLM lifecycle, with the final implication addressing cross-cutting issues such as documentation and tooling, as shown in Fig.~\ref{fig:Implications}. The blue box presents the steps of how to conclude implications from the findings in RQ1 and RQ2, the red box summarizes the limitations of the existing DL frameworks and their LLM toolkits in supporting LLMs, while the green box refers to the relevant findings, and the purple box presents the optimization suggestions for each limitation.\looseness=-1

\faExclamationTriangle\ \textbf{Environment Setup Fragility and Environment Reproducibility Gaps.} 
This challenge is derived from Findings 1, 4, and 5, related to the environment setup stage of the whole LLM development lifecycle. Specifically, we find that a consistent theme across the findings is the fragility of the environment setup in LLM development. Finding 1 identifies installation and runtime errors (27.31\%, A.1) as the most frequently reported user questions, typically caused by vague setup instructions, dependency mismatches, and incompatible CUDA/cuDNN versions. Finding 4 highlights how newcomers are frequently blocked at the very first step due to cryptic errors and insufficient onboarding support. Complementing these, Finding 5 confirms that developers frequently observe deployment failures rooted in rigid toolchains, fragile integration logic, and hardware mismatches.\looseness=-1

This challenge is particularly acute during the initial environment setup phase, where GPU drivers, compiler toolchains, Python dependencies, and accelerator-specific packages (e.g., apex for MindSpeed) must align precisely. Minor discrepancies often lead to cascading errors that are difficult to trace and reproduce. Unfortunately, DL frameworks seldom provide robust abstraction layers or compatibility validation mechanisms, leaving users to manually diagnose environment-related failures. Even experienced LLM users struggle to reproduce environments across machines, and cross-platform compatibility remains unreliable without containerization or virtualization, which introduces additional complexity.\looseness=-1

\textbf{Threats on LLM Users.} Environment setup failures directly hinder project initiation and experimentation. New users are frequently blocked by low-level compatibility errors that they lack the expertise to resolve. Experienced developers face high overhead in maintaining reproducible environments across systems, slowing iteration cycles and discouraging long-term adoption of the framework.\looseness=-1

\faExclamationTriangle\ \textbf{Missing Cohesive Abstractions for LLM Workflows.} 
This challenge is derived from Findings 1, 2, and 4, related to the model development stage of the whole LLM development lifecycle. Specifically, we find that these findings collectively reveal that model development workflows, spanning checkpointing, precision handling, and parameter control, lack sufficient abstraction and integration support. Finding 1 reports frequent user confusion over training configurations (15.31\%, A.2 and A.5), especially in multi-GPU settings or when customizing optimizer schedules. Finding 2 shows that 36.42\% of user requirements (B.2) call for modular APIs, distributed training scalability, and format interoperability. Furthermore, Finding 4 emphasizes that even expert users suffer from undocumented behaviors, implicit design assumptions, and a lack of integrated workflows.\looseness=-1

These observations indicate that while low-level primitives are exposed by most frameworks, high-level, cohesive abstractions remain missing. Tasks such as enabling parameter-efficient fine-tuning (e.g., LoRA), managing eval/train modes, or freezing model subcomponents are scattered across different modules, with limited documentation or examples. Inconsistent API behavior across components frequently leads to incomplete checkpoints, unstable training states, and misaligned outputs. Moreover, implicit behaviors, such as dropout or normalization effects under different modes, are rarely surfaced to the user, exacerbating silent runtime failures and increasing the cognitive burden of debugging.\looseness=-1

\textbf{Threats on LLM Users.} The lack of structured development support forces users into ad hoc, trial-and-error practices. Debugging becomes time-consuming, and silent misconfigurations can corrupt model correctness without warning. This not only reduces development efficiency but also erodes user confidence in the correctness and reproducibility of their results.\looseness=-1

\faExclamationTriangle\ \textbf{Behavioral Instability and Execution Non-Determinism.} 
This challenge is derived from Findings 1, 3, 4, and 5, related to the execution optimization and behavioral stability stage of the whole LLM development lifecycle. Specifically, we find that the instability and poor reproducibility of execution behaviours are recurrent concerns across both user and developer perspectives. Finding 1 identifies unexpected behaviors, such as abnormal loss, repeated outputs, or unstable gradients, as the largest category of user-reported questions (30.07\%, A.3), often lacking diagnostic clarity. Finding 3 confirms that numerical instability, architectural inconsistencies, and flawed control logic (C.2, C.4, C.5, C.8) are common causes of execution-stage bugs. Additionally, Findings 4 and 5 reveal that both users and developers face silent failures, non-deterministic behaviors, and state inconsistencies during training and inference.\looseness=-1

These issues manifest during the core execution phase, where runtime safeguards are notably absent. Forgotten model.eval() switches, dropout misconfiguration, or silent precision drift under mixed-precision training often produce corrupted results without error messages. In complex setups like ZeRO3 or LoRA, checkpoint restoration frequently leads to misaligned weights or diverging loss patterns. Inference bugs, such as token repetition or attention misalignment, further compound reproducibility challenges. Despite the severity of these issues, frameworks lack runtime assertions, behavioral validators, or reliable logging utilities to surface such failures early in the workflow.

\textbf{Threats on LLM Users.} Instability during execution undermines trust in model behavior. Silent errors compromise experiment integrity, delay fault localization, and obstruct reproducibility, critical for both research replication and production deployment. In the absence of stable execution guarantees, users cannot confidently scale or iterate on their models.\looseness=-1

\faExclamationTriangle\ \textbf{Inefficient Resource Utilization and Profiling Blind Spots.}
This challenge is derived from Findings 3, 4, and 5, related to the execution optimization and system efficiency stage of the whole LLM development lifecycle. Specifically, we find that performance bottlenecks related to memory scheduling, parallelism configuration, and profiling limitations are widely reported in both issue analyses and interviews. Finding 3 highlights frequent bugs associated with resource inefficiency and distributed execution (C.2, C.3), including scheduling stalls and GPU underutilization. Finding 4 mentions user frustrations with invisible regressions caused by fused kernel mismatches and recomputation logic. Finding 5 further underscores that kernel dispatch delays, backend inconsistencies, and lack of fault isolation impede runtime efficiency.\looseness=-1

During large-scale training, these issues arise from both software inefficiencies and poor observability. Suboptimal memory allocation, imprecise recomputation policies, and fallback kernels (e.g., in FlashAttention) degrade operator performance. Distributed strategies like tensor, pipeline, or sequence parallelism are often fragile, with unclear configuration interfaces and limited fault tolerance. Frameworks typically offer insufficient support for runtime profiling or visualization, making it difficult to attribute bottlenecks to specific stages or components.\looseness=-1

\textbf{Threats on LLM Users.} System inefficiencies prolong training time, waste computing resources, and inflate experimentation costs. In high-scale deployments, such inefficiencies become a major barrier to iteration, reproducibility, and deployment readiness, especially for users with limited hardware or budget.\looseness=-1

\faExclamationTriangle\ \textbf{Observability, Documentation, and Contribution Workflow Gaps.}
This challenge is derived from Findings 2, 3, and 5, related to all the stages of the LLM development lifecycle. Specifically, we find that tooling and documentation deficiencies pervade every stage of the LLM development lifecycle. Finding 2 notes that 7.89\% of requirements (B.4) request better onboarding support, tutorials, and collaboration tools. Finding 3 reveals that 39.29\% of bugs (C.6–C.10) involve maintainability or observability failures, including broken logging, missing checkpoints, and unclear error messages. From the developer side, Finding 5 underscores that poor documentation, fragmented workflows, and a lack of reusable tools make framework maintenance and collaboration difficult.\looseness=-1

These gaps manifest in various forms: inconsistent documentation, outdated tutorials, missing type hints, and fragile testing infrastructure. Users frequently struggle to locate accurate references or examples that match current API behavior. Tools for visualization, debugging, or performance tracking are either non-existent or require deep customization, further deterring adoption. Moreover, frameworks often provide vague contribution templates and lack standards for reproducibility or testing, discouraging community participation and slowing ecosystem evolution.\looseness=-1

\textbf{Threats on LLM Users.} The absence of accessible tooling and reliable documentation imposes friction at every stage from installation to debugging and contribution. This increases onboarding time, hinders productivity, and limits framework extensibility. Without first-class developer experience, even technically capable frameworks fail to gain sustained community support.\looseness=-1

\textbf{Conclusion.} These five findings correspond to the key stages of LLM development and collectively reveal how current DL frameworks fall short in supporting real-world LLM development lifecycle. Finding 1 concerns environmental compatibility, which directly affects users’ ability to begin development. Without a stable hardware–software stack, progress halts at the outset. Once the setup completes, Finding 2 reflects challenges in using core functionalities, where poor abstraction and weak documentation lead to frequent misconfigurations. As users move into training, Finding 3 highlights instability and irreproducibility caused by incorrect state transitions or silent control-logic errors. Even when training succeeds, Finding 4 reveals performance bottlenecks such as inefficient scheduling and underutilized hardware that limit scalability. Finally, Finding 5 shows that lacking ecosystem support, including tooling, documentation, and collaborative workflows, slows down onboarding and hinders long-term maintenance.\looseness=-1

\vspace{-4mm}
\subsection{Optimizations for DL Frameworks in Supporting LLMs}

To address the challenges identified in the previous section, we propose five optimization directions grounded in empirical evidence and interview feedback. These recommendations aim to improve framework usability, reliability, and performance across the LLM lifecycle. Each targets a specific pain point observed in real-world usage and contributes practical, actionable improvements.\looseness=-1

\faMagic\ \textbf{Simplify Environment Setup and Improve Reproducibility.}
To lower the entry barrier and improve deployment robustness, we recommend simplifying the environment setup process through standardized and validated configuration workflows. Frameworks should offer hardware-specific deployment kits, including prebuilt Docker containers, conda environment files, and installation scripts that cover common accelerator backends (e.g., CUDA, ROCm, Ascend). These kits should automatically verify driver compatibility, resolve common dependency mismatches (e.g., apex, flash-attention), and surface actionable error messages. Additionally, guided onboarding tutorials should demonstrate reproducible workflows for common tasks, such as fine-tuning BERT or deploying LLaMA models on heterogeneous hardware.\looseness=-1

This optimization directly addresses the challenge identified in Limitation 1, where users are frequently blocked by opaque errors, version mismatches, and fragile toolchain dependencies. By automating compatibility checks and abstracting low-level configuration concerns, this approach minimizes manual intervention and reduces the cognitive burden on users, particularly in GPU/NPU-centric environments. Verified onboarding pipelines also improve reproducibility across platforms, a recurring pain point for both end users and developers.\looseness=-1

\textbf{Implications for Framework Developers.} Developers should treat environment setup as a core part of framework design rather than an external concern. Providing hardware-aware setup tools and reproducible deployment recipes enables more reliable experimentation, facilitates onboarding, and encourages adoption, especially in scenarios involving novel accelerators or institutional clusters.\looseness=-1

\faMagic\ \textbf{Provide Cohesive High-Level Abstractions for LLM Workflows.}
To facilitate modular and error-resistant model development, frameworks should provide structured APIs for common LLM functionalities, such as checkpointing, precision switching, parameter freezing, and training/evaluation mode transitions. These APIs should be exposed through a unified configuration interface that orchestrates module behavior across distributed devices. Additionally, reference configurations for popular LLMs (e.g., LLaMA, Yi, Mistral) should be included to guide users in assembling scalable and robust training pipelines.\looseness=-1

This recommendation directly mitigates the fragmentation observed in Limitation 2, where users struggle to coordinate low-level operations due to inconsistent interfaces and implicit behaviors. By promoting composable abstractions for critical tasks, such as optimizer freezing or mixed-precision setup, frameworks can reduce silent configuration errors and streamline multi-component model workflows. These abstractions also serve as a foundation for extensibility, making advanced methods like LoRA and ZeRO3 easier to adopt.\looseness=-1

\textbf{Implications for Framework Developers.} Rather than exposing only low-level primitives, framework developers should elevate common LLM operations into well-defined, high-level modules. This shift enables safer composition, better reuse, and more transparent behavior, improving both usability and maintainability in large-scale model development.\looseness=-1

\faMagic\ \textbf{Safeguard Execution with Runtime Consistency Checks.}
To detect silent failures and improve runtime robustness, we recommend integrating consistency checks into training and inference workflows. Frameworks should validate runtime conditions such as eval() mode before inference, detect NaNs or divergence patterns in loss trajectories, enforce seed propagation across distributed workers, and verify checkpoint compatibility before loading. These checks should be lightweight but informative, surfacing hidden logic errors before they corrupt training results or degrade model quality.\looseness=-1

This optimization directly addresses the issues in Limitation 3, where execution-phase instability and non-determinism lead to corrupted states, diverging losses, and silent runtime failures. By instrumenting guardrails at key transition points, such as training-to-inference or checkpoint-to-resume, frameworks can catch subtle errors early, reducing the need for time-consuming manual debugging and improving reproducibility guarantees.\looseness=-1

\textbf{Implications for Framework Developers.} Execution correctness must be actively safeguarded. Embedding runtime assertions and safety checks can prevent small misconfigurations from escalating into untraceable bugs, especially in large-scale or mixed-precision settings. These investments enhance user trust and reduce the cost of failure in complex training pipelines.\looseness=-1

\faMagic\ \textbf{Enable Lightweight Observability for Performance Tuning.}
To address runtime inefficiencies, frameworks should embed lightweight observability tools that provide users with actionable performance insights. These include memory profilers, kernel activity trackers, and communication monitors that visualize operator-wise time consumption and resource utilization. Logs should flag fallback execution paths, such as disabled FlashAttention kernels, and provide suggestions for remediation. In addition, cost estimation utilities should help users compare the resource footprint of different configurations to guide scalable training decisions.\looseness=-1

This optimization is targeted at the bottlenecks identified in Limitation 4, where misconfigured parallelism and inefficient resource usage frequently go unnoticed due to a lack of feedback mechanisms. By exposing runtime metrics and diagnostics in an interpretable manner, users can reason about inefficiencies, fine-tune their setups, and avoid unnecessary compute waste. Importantly, such visibility helps bridge the knowledge gap for users without deep systems expertise.\looseness=-1

\textbf{Implications for Framework Developers.}Transparency is key to performance engineering. Rather than optimizing for raw throughput alone, DL frameworks should provide interpretability and feedback, empowering users to make informed trade-offs and self-diagnose performance regressions without deep internal knowledge.\looseness=-1

\faMagic\ \textbf{Strengthen Documentation, Tooling, and Collaboration Workflows.}
To improve the overall developer experience and reduce onboarding cost, frameworks should treat documentation and tooling as integral components. Documentation should adopt a task-oriented format, combining runnable scripts, annotated configurations, expected outputs, and environment specifications. Tutorials should walk through complete workflows (e.g., training LLaMA with LoRA and ZeRO3) and be automatically tested for consistency with current APIs. Developer tools such as profilers, activation loggers, and tensor visualizers should be standardized and distributed as reusable plugins. Community collaboration should be supported through structured templates, contribution guides, and module ownership definitions.\looseness=-1

This recommendation responds to the pervasive infrastructure issues described in Limitation 5, where users face outdated documentation, broken examples, and missing tools. By strengthening the surrounding ecosystem, through better onboarding support, standardized toolchains, and reproducibility infrastructure, frameworks can ensure consistent usage, reduce developer friction, and foster an active contributor base.\looseness=-1

\textbf{Implications for Framework Developers.} Developer support infrastructure is not a peripheral concern; it is central to the long-term health of a framework ecosystem. Investing in modular documentation, validated tutorials, and transparent collaboration workflows accelerates user adoption, reduces technical debt, and improves the maintainability of LLM tooling at scale.\looseness=-1

Overall, these measures address environment fragility, missing functionality, silent failures, performance inefficiencies, and ecosystem gaps. They offer a practical roadmap for improving DL frameworks to meet the demands of real-world LLM development.\looseness=-1

\vspace{-2mm}
\section{Threats to Validity}

\textbf{Internal Validity.} This threat concerns the reliability of the filtered issue reports and the consistency of interview interpretation. First, selecting LLM-centric issues depends on a predefined set of filter tags, which may cause misclassification if applied inconsistently. To mitigate this, all annotators manually verify the relevance of each candidate tag and reach consensus before applying it, ensuring that the selected tags consistently reflect LLM-centric content. Second, interview interpretation may be affected by selective emphasis or omission of certain remarks during analysis. To address this, the first and second authors independently reviewed the full transcripts, resolved discrepancies through discussion, and reached a final agreement to reduce interpretation bias. Third, the interviewer–interviewee interaction may influence the depth and direction of responses. To reduce this risk, we followed an interview with open-ended questions and minimized intervention during participants’ initial responses to allow spontaneous expression.\looseness=-1

\textbf{External Threats.} This threat concerns the generalizability of our findings, particularly regarding the diversity of interviewees and the representativeness of the selected DL frameworks and toolkits. To address the first concern, we include both framework developers and LLM users with varying levels of experience. Most LLM users are students or early-career researchers engaged in academic-scale development. While their perspectives may not fully capture the operational complexity of production-grade systems, they offer valuable insights into the usability, accessibility, and learning curve of LLM frameworks, factors critical for broader adoption. These developers constitute a substantial portion of the open-source and research-driven LLM community, making their feedback highly relevant. To improve generalizability, future work will involve more interviewees from industrial environments, especially those with hands-on experience in deploying and maintaining large-scale systems. Collaborating with open-source communities and companies may further enhance ecological validity.
The second concern relates to the representativeness of the frameworks and toolkits included in our study. We collect issue reports from three widely used DL frameworks that support LLMs and from eight representative LLM toolkits. These toolkits include widely adopted industrial solutions (e.g., DeepSpeed, Megatron), emerging high-performance libraries (e.g., vLLM, TensorRT-LLM), and rapidly evolving ecosystems (e.g., MindFormers, MindSpeed). All issue reports are sampled from 2023 onward to ensure both timeliness and relevance. In future work, we plan to expand the scope to include additional frameworks and assess the generalizability of our taxonomy across a broader range of LLM development platforms.\looseness=-1

\textbf{Construct Threats.} This threat concerns whether our methods accurately capture the intended constructs, particularly the taxonomy of LLM-centric framework challenges and the themes derived from interviews. We identify three key risks to construct validity.
First, our keyword-based filtering approach may affect the accuracy of post selection. Since identification of LLM-centric posts relies on predefined keywords, the filter may omit relevant discussions or include unrelated ones. We address this by manually reviewing a sample of filtered posts, iteratively refining the keyword set, and ensuring alignment with the intended scope.
Second, the initial round of open coding was conducted by a single researcher, which may introduce subjective bias into early interpretations. Although final themes were refined through collaborative discussions, initial framing decisions may still influence the outcome. We mitigate this by documenting all coding rationales in detailed memos, grounding codes in direct quotes, and maintaining a transparent, traceable analysis process. In future work, we plan to involve multiple coders and assess inter-coder agreement using metrics such as Cohen’s Kappa.
Third, our two-stage labeling process may reduce coder independence. One researcher conducted the pilot coding and produced an initial taxonomy, which subsequent annotators used as a starting point. This structure may anchor later judgments and suppress alternative categorizations. To reduce this risk, we encourage re-evaluation during team discussions and record disagreement cases. Final inter-coder agreement scores exceed 0.7, suggesting acceptable reliability.\looseness=-1

\vspace{-2mm}
\section{Related Work}
\label{sec:relatedwork}

Developers and users of DL frameworks often discuss issues when using frameworks and those exposed bugs in the open source community, such as GitHub and StackOverflow. Therefore, researchers often conduct empirical studies and mine the technical posts and bug reports to analyze and conclude the technical challenges and bug-fixing patterns. Therefore, such work can be divided into (1) studies on the application of DL frameworks and (2) studies on DL framework bugs.\looseness=-1

\textbf{Studies on the Application of DL Frameworks.}
Nguyen et al.~\cite{nguyen2019machine} surveyed machine learning (ML) and deep learning (DL) trends, examining scalability challenges and the dynamic big data ecosystems. The study identified seven crucial insights for tool selection aimed at optimized performance and real-world problem resolution.
Han et al.~\cite{han2020programmers} analyzed the discussion posts about three popular DL frameworks (i.e., TensorFlow~\cite{tensorflow}, PyTorch~\cite{torch}, and Theano~\cite{Theano}) on Stack Overflow and GitHub. They identified critical trends and discrepancies, predominantly between TensorFlow and PyTorch, offering constructive insights to developers, users, and researchers for improved DL framework utilization and development.
Wang et al.~\cite{wang2021automatic} conducted the maiden empirical study on unit test generation techniques for ML libraries. They revealed that existing libraries often possess unit test suites of low value, while other tools such as EVOSUITE~\cite{fraser2011evosuite} and Randoop~\cite{pacheco2007randoop,pacheco2007feedback} can enhance code coverage and mutation score of mutation testing~\cite{mutate1971,DeMillo1978HintsOT}. \looseness=-1

\textbf{Studies on DL Framework bugs.} Bugs in DL frameworks often trouble both developers and users. Commonly, users report issues found during development to the relevant open-source communities for resolution. These reports, frequently including user-submitted bug reports and technical discussions, have been the focus of extensive research to identify prevalent bug patterns, understand user challenges, and glean insights for improving DL frameworks.\looseness=-1

Some researchers collect bug reports, analyze their characteristics, and identify feature patterns to understand the root causes of framework bugs. Du et al.\cite{du2022empirical} introduce a fault-trigger perspective by distinguishing between Bohrbugs that can be consistently reproducible and Mandelbugs which cannot be reproduced even under the same condition), emphasizing the link between reproducibility and testing priority. Zhang et al.~\cite{zhang2018empirical} study TensorFlow-related bugs by manually analyzing 175 reports from GitHub and Stack Overflow. Zhang et al.~\cite{zhang2020empirical} investigate 4,960 DL job failures from Microsoft’s Philly system and manually sample 100 cases for in-depth analysis.
Chen et al.\cite{chen2023toward} examine 800 bugs from TensorFlow, PyTorch, and MXNet to extract shared features and inform repair strategies. Li et al.~\cite{jia2021symptoms} analyze TensorFlow bug reports to reveal distribution patterns and support efficient detection and localization. Han et al. categorize user requirements based on GitHub and Stack Overflow data, highlighting developers’ pain points with DL frameworks.
Other studies replicate triggering conditions to assess the behavior and impact of specific bug types. Tambon et al.\cite{tambon2024silent} detect ``silent'' bugs in Keras~\cite{chollet2015keras} and TensorFlow~~\cite{tensorflow} that fail silently without error messages or crashes, offering insights for proactive prevention. Hong et al.\cite{hong2024investigating} address similar bugs in PyTorch using LLMs and propose the PYSIASSIST tool for automated debugging. Makkouk et al.~\cite{makkouk2022empirical} investigate performance bugs in PyTorch and TensorFlow that degrade real-time feedback, analyzing their computational cost and suggesting mitigation strategies. Zhang et al.~~\cite{zhang2020empirical} further explore framework bugs that cause long-running application crashes and propose practical repair techniques.\looseness=-1

\renewcommand{\arraystretch}{1.4} 
\begin{table}[]
  \centering
  \vspace{-4mm}
  \caption{Comparison between Previous Work and Our Study}
  \vspace{-4mm}
  \resizebox{1.02\textwidth}{!}{
    \begin{tabular}{>{\centering\arraybackslash}m{0.25\textwidth}|l}
    \hline
    Comparison Dimension & \multirow[c]{1}{*}{Description} \\
    \hline
    \multirow[c]{2}{*}{\parbox[t]{0.25\textwidth}{\centering Scope of Analysis}}
        & \ding{117} Focuses on general-purpose models (e.g., CNNs, RNNs) and frameworks like TensorFlow or PyTorch \\
        & \ding{51} Focuses on frameworks and toolkits used for large language models (LLMs), such as Megatron and DeepSpeed \\
    \hline
    \multirow[c]{2}{*}{\parbox[t]{0.25\textwidth}{\centering Coverage of Taxonomy}}
        & \ding{117} Analyzes system-level runtime errors and bug symptoms, covers bugs only, ignoring usability issues and missing features \\
        & \ding{51} Classifies issues as Question, Requirement, and Bug to capture full-spectrum framework challenges \\
    \hline
    \multirow[c]{2}{*}{\parbox[t]{0.25\textwidth}{\centering Granularity of\\ Classification}}
        & \ding{117} Uses coarse-grained symptom types like crash or memory error \\
        & \ding{51} Introduces 3 types, 20 themes, and over 75 sub-themes across the LLM development lifecycle \\
    \hline
    \multirow[c]{2}{*}{\parbox[t]{0.25\textwidth}{\centering Development Stage\\ Mapping}}
        & \ding{117} Typically focuses on runtime bugs only \\
        & \ding{51} Explicitly links each issue type to a development stage (installation, training, deployment, etc.) \\
    \hline
    \multirow[c]{2}{*}{\parbox[t]{0.25\textwidth}{\centering Problem Presentation\\ Levels}}
        & \ding{117} Focuses on crashes, exceptions, and incorrect outputs \\
        & \ding{51} Captures functional gaps (B), user confusion (A), and silent failures (C) \\
    \hline
    \multirow[c]{2}{*}{\parbox[t]{0.25\textwidth}{\centering Problem Frequency under\\ LLM Settings}}
        & \ding{117} Some types appear infrequently under small models or CV/NLP tasks \\
        & \ding{51} LLMs trigger these types frequently due to system scale and coupling \\
    \hline
    \multirow[c]{2}{*}{\parbox[t]{0.25\textwidth}{\centering Root Cause\\ Modeling}}
        & \ding{117} Rooted in static symptoms or exception types \\
        & \ding{51} Links issues to triggering context, system mechanisms, and user intent \\
    \hline
    \multirow[c]{2}{*}{\parbox[t]{0.25\textwidth}{\centering Debugging and Repair\\ Guidance}}
        & \ding{117} Propose concrete repair patterns, detection rules, or patching strategies for system-level bugs \\
        & \ding{51} Models error-trigger-distress-expectation chains to aid diagnosis, but lacks concrete repair methods \\
    \hline
    \multirow[c]{2}{*}{\parbox[t]{0.25\textwidth}{\centering Validation\\ Method}}
        & \ding{117} Based solely on GitHub issue mining \\
        & \ding{51} Validated and iteratively refined through interviews with 11 LLM users and 8 framework developers \\
    \hline
    \multirow[c]{2}{*}{\parbox[t]{0.25\textwidth}{\centering Goal\\ Orientation}}
        & \ding{117} Aims at reliability analysis or bug testing \\
        & \ding{51} Targets framework usability, configuration robustness, and developer experience \\
    \hline
    \multirow[c]{2}{*}{\parbox[t]{0.25\textwidth}{\centering Consideration of Documentation\\ and Community}}
        & \ding{117} Often excludes doc errors or contributor experience \\
        & \ding{51} Captures documentation drift (C.10) and community support gaps (B.4) \\
    \hline
    \end{tabular}%
  }
  \vspace{1ex}
  \footnotesize{\textit{Note:} \ding{117} = Previous Work, \ding{51} = Our Work.}
  \label{tab:relatedwork}%
\end{table}

\textbf{Novelty of Our Study.} Compared to previous studies, we adopt a ``symptom × root cause'' perspective, analyze real-world bug reports from framework communities such as GitHub and Gitee, and emphasize a structured, multi-dimensional classification scheme. Our taxonomy introduces several key differences that render existing classifications insufficient for capturing challenges in LLM-centric DL frameworks.
LLM development introduces distinct stressors at every stage. Environment setup becomes more fragile due to tighter hardware–toolchain coupling. Core functionalities demand precise coordination of mechanisms such as parameter freezing and mixed-precision scaling. Execution errors are often silent yet highly disruptive, involving numerical drift and control-state inconsistency that are difficult to detect. Performance tuning grows critical under extreme computational scales. Usability degrades rapidly as evolving toolchains outpace documentation and ecosystem support.
These problems are not scaled-up versions of previous issues but differ qualitatively in frequency, impact, and diagnosability. Addressing them requires treating LLM support as a distinct systems engineering problem that calls for dedicated abstractions, robust defaults, and lifecycle-aware tooling.  Next, we list three key differences between previous studies and us.\looseness=-1

First, previous studies focus mainly on defect reports and overlook challenges reflected in user questions and requirements. Although previous research systematically analyzes defect types in DL frameworks, most restrict their scope to traditional bug reports. In real-world LLM development, however, developers frequently raise not only functional failures but also feature requests and confusion-driven questions. These non-defect reports, such as hardware support needs, training misconfigurations, or behavioral changes after version upgrades, do not represent explicit bugs but reveal flaws in framework design, documentation, and usability. To address this gap, our taxonomy introduces two additional categories: user questions, reflecting knowledge gaps, and user requirements, reflecting functionality gaps. This broader perspective captures a more complete range of challenges in LLM development and complements previous defect-centric studies. \looseness=-1

Second, existing taxonomies target small and medium-scale models and fail to capture LLM-specific defect patterns. Most previous studies focus on conventional models such as ResNet, VGG, or BERT-base, which are relatively small, structurally simple, and run in stable, single-device settings. As a result, they mainly address common issues such as API misuse, type mismatches, or build errors. In contrast, LLMs introduce distinct complexities during training and inference. Hybrid parallelism (tensor, pipeline, and expert parallelism) in LLM training is highly sensitive to misconfigurations, often leading to synchronization errors or deadlocks. Modular fine-tuning methods such as LoRA or Adapter demand precise control over optimizer separation, parameter freezing, and gradient flow, making them prone to failure. During inference, aggressive quantization enhances efficiency but increases the risk of numerical instability, which may cause degraded outputs such as repeated tokens or incoherent text. These issues seldom arise in smaller models or have a limited impact when they do. Therefore, existing taxonomies lack the granularity to describe defects rooted in the scale, architectural complexity, and execution diversity of LLMs. To address this gap, we analyze a large set of LLM-centric issue reports and construct a taxonomy with 10 top-level categories and over 40 subtypes, systematically covering defects in parallelism, numerical precision, and inference behavior.\looseness=-1

Third, previous work lacks validation from real LLM users and framework developers.
Most studies rely on static sources such as GitHub issues, Stack Overflow posts, or internal logs, which researchers manually categorize. While this approach reveals common problems, it misses subtle but critical issues like strategy degradation or silent configuration failures. Moreover, existing taxonomies are often researcher-defined and do not reflect how developers think or debug. To address this, we construct our taxonomy and then validate it through interviews with 11 LLM users and eight framework developers. Their feedback clarifies category boundaries and prompts revisions to several subcategories. This validation ensures our taxonomy aligns with real-world practices and captures challenges overlooked by previous work.\looseness=-1

In summary, previous studies provide a useful foundation for understanding DL framework defects but remain limited in issue diversity, model scale, and empirical depth. Our work addresses these gaps through a tripartite classification covering questions, defects, and requirements, with a focus on LLM-specific challenges and validation grounded in large-scale community data and real developer feedback. More fine-grained comparison between previous work and our study is shown in Table~\ref{tab:relatedwork}. Each row represents a specific comparison dimension, covering aspects such as analytical scope, classification coverage, granularity, lifecycle mapping, root cause modeling, and validation methodology. The second column summarizes common characteristics of existing studies, while the third column highlights how our work differs or advances in each regard.\looseness=-1

\section{Conclusion}
The rapid advancement of LLMs places growing demands on DL frameworks, which serve as their fundamental infrastructure. To investigate the challenges DL frameworks face in supporting LLMs, we conduct a comprehensive study that combines large-scale issue report analysis from three major DL frameworks and eight LLM toolkits with in-depth interviews of LLM users and framework developers. Based on this empirical analysis, we construct a taxonomy covering LLM-centric bugs, user questions, and requirements. We then identify five key findings that expose critical limitations in current frameworks. Building on these insights, we propose five actionable optimization strategies to guide future framework design and testing practices.

\section{Data Availability}
\label{sec:dataavailable}
All details of the taxonomy and interviews are available on our project homepage~\cite{sharelink}.

\section{Acknowledgement}
\label{sec:Acknowledgement}
This work is partially supported by the National Key Research and Development Program of China (2024YFF0908001) and the National Natural Science Foundation of China (U24A20337 and 62372228), the Shenzhen-Hong Kong-Macau Technology Research Programme (Type C) (Grant No.SGDX20230821091559018), and the Fundamental Research Funds for the Central Universities (14380029).

\bibliographystyle{IEEEtran}
\bibliography{sample-base}

\end{document}